\documentclass[paper,notoc]{JHEP3}

\usepackage{epsfig}
\usepackage{amsmath}
\usepackage{amssymb}
\usepackage{graphicx}

\usepackage{verbatim}   % for \begin{comment} \end{comment}
\usepackage{multirow}   % for multirow's in tables

\allowdisplaybreaks[1]

\DeclareGraphicsExtensions{eps,ps}

\def \refeq#1{(\ref{#1})}  
\def \refsec#1{Section \ref{#1}}
\def \refSec#1{Section \ref{#1}}
\def \refapp#1{Appendix \ref{#1}}
\def \reffig#1{Figure \ref{#1}}
\def \reftab#1{Table \ref{#1}}

\def \nn{\nonumber\\}
\def \dis{\displaystyle}

\def \order#1{ {\cal O} \left( #1 \right) }

\def \cM{{\cal M}}

\def \cLdB1{{{\cal L}_{\Delta B = 1}^{\rm EW}}} % Delta_B = 1 effective Lagrangian
\def \BR{{\cal B}}                               % branching ratio 

\def \Op{{\cal O}}

\def \One{\leavevmode\hbox{\small1\kern-3.6pt\normalsize1}} % unit matrix

\def \MeV{{\rm \; MeV}}
\def \GeV{{\rm \; GeV}}

\def \gS{g_s}              % strong coupling
\def \alS{\alpha_s}        % strong coupling
\def \alE{\alpha_e}        % electro-magnetic coupling
\def \GF{G_F}              % Fermi coupling
\def \LamConf{{\Lambda_{\rm QCD}}}    %  confinement scale
\def \Gaml{{\Gamma_l}}
\def \Game{{\Gamma_e}}
\def \Gammu{{\Gamma_\mu}}

\def \BRl{{{\cal B}_l}}
\def \BRe{{{\cal B}_e}}
\def \BRmu{{{\cal B}_\mu}}

\def \AFBl{{A_{\rm FB}^l}}
\def \AFBmu{{A_{\rm FB}^\mu}}
\def \AFBe{{A_{\rm FB}^e}}

\def \FHl{{F_H^l}}
\def \FHe{{F_H^e}}
\def \FHmu{{F_H^\mu}}

\def \BXclv  {\bar{B} \to X_c l \bar\nu_l}

\def \bsll {b\to s \bar{l}l}

\def \BRincl#1#2#3{{\BR^{\rm incl}_#1 |_{[#2] #3}}}

\def \BtoKll{{\bar{B} \to K \bar{l}l}}
\def \BtoKee{{\bar{B} \to K \bar{e}e}}
\def \BtoKmumu{{\bar{B} \to K \bar{\mu}\mu}}

\def \BmtoKmll{{B^- \to K^- \bar{l}l}}
\def \BneutralKll{{\bar{B}^0 \to K^0 \bar{l}l}}
\def \BtoKastll{{\bar{B} \to K^\ast \bar{l}l}}

\def \Bstoll{{\bar{B}_s \to \bar{l}l}}
\def \Bstomm{{\bar{B}_s \to \bar{\mu}\mu}}

\def \Bstoee{{\bar{B}_s \to \bar{e}e}}

\def \BmtoKmmumu{{ B^- \to K^- \bar{\mu} \mu}}
\def \B0toK0mumu{{ \bar{B}^0 \to K^0 \bar{\mu} \mu}}
\def \BXsll{\bar{B} \to X_s \bar{l}l}

\def\MSbar{{\overline{\rm MS}}}    % MS-bar

\def \CF{{C_F}}    % 
\def \Nc{{N_c}}    % number of colors

\def\D0{D\O}  \def\d0{D\O}
\title{Angular Distributions of $\BtoKll$ Decays}

\author{Christoph Bobeth, Gudrun Hiller and Giorgi Piranishvili \\
  Institut f{\"u}r Physik, Technische Universit{\"a}t Dortmund, D-44221
  Dortmund, Germany}
\date{\today}

\abstract{We model-independently analyze the angular distributions of $\BtoKll$
  decays, $l=e, \mu$, for low dilepton mass using QCD factorization.  Besides
  the decay rate, we study the forward-backward asymmetry $\AFBl$ and a further
  observable, $\FHl$, which gives rise to a flat term in the angular
  distribution. We find that in the Standard Model $\FHl \propto m_l^2$, hence
  vanishing $\FHe$ and $\FHmu$ of around $2 \%$ (exact value depends on cuts)
  with a very small theoretical uncertainty of a few percent. We also give
  predictions for $R_K$, the ratio of $\BtoKmumu$ to $\BtoKee$ decay rates. We
  analytically show using large recoil symmetry relations that in the Standard
  Model $R_K$ equals one up to lepton mass corrections of the order $10^{-4}$
  including $\alpha_s$ and subleading $1/E$ power corrections.
  The New Physics reach of the observables from the $\BtoKll$ angular analysis
  is explored together with $R_K$ and the $\bar B_s \to \bar l l$ and $\BXsll$
  branching ratios for both $l=e$ and $l=\mu$. We find substantial room for
  signals from (pseudo-) scalar and tensor interactions beyond the Standard
  Model. Experimental investigations of the $\BtoKmumu$ angular distributions
  are suitable for the LHC environment and high luminosity B factories, where
  also studies of the electron modes are promising.}

\keywords{B-Physics, Beyond Standard Model, Rare Decays}

\preprint{DO-TH 07/07}

\begin{document}

%
%
%----------------------------------------------------
\section{Introduction}

The exclusive decays $\BtoKll$ with $l= e, \mu$ are governed in the Standard
Model (SM) by flavor-changing neutral currents, and hence constitute sensitive
probes of New Physics (NP). The three-body decays allow to study non-trivial
observables by kinematical measurements of the decay products.  They give access
to a double differential decay spectrum with respect to the invariant mass of
the lepton pair $q^2$ and a lepton charge asymmetry angle $\cos\theta$. In the
absence of large statistics, partially integrated spectra such as the dilepton
mass spectrum $d\Gaml/dq^2$ or the angular distribution $d\Gaml/d\!\cos\theta$
can be explored. Further $\Gaml \equiv \Gamma(\BtoKll)$ is in general different
for electrons and muons. Having various theoretical or experimental advantages,
the $\BtoKll$ observables cover a wide range of SM tests and NP searches, that
are well suited for experimental study at high luminosity facilities at the
$\Upsilon(4S)$ and the Large Hadron Collider (LHC), e.g.,~\cite{superb}.

The $\BtoKll$ branching ratio has been determined experimentally to be in
agreement with the SM within uncertainties, and lies in the $10^{-7}$ region
\cite{Abe:2004ir, Ishikawa:2006fh, Aubert:2006vb, Scuri:2007py,Ali:1999mm}.
Early data on more elaborate observables and $q^2$-spectra are beginning to come
from the B factories \cite{Abe:2004ir, Ishikawa:2006fh,Aubert:2006vb}.  While
theoretical studies presented extensive phenomenological analyses of the
dilepton mass distribution \cite{Ali:1999mm, Bobeth:2001sq}, a detailed
exploration of the SM background and NP potential of the angular dependence in
the decay distribution is lacking. The $\BtoKll$ angular distribution is very
simple in the SM \cite{Ali:1999mm, Bobeth:2001sq}
\begin{equation}
  \label{eq:angular:SM}
   \frac{d\Gaml^{\rm SM}}{d\!\cos\theta}  \propto 
   \sin^2\theta + {\cal{O}}(m_l^2),
\end{equation}
up to small lepton mass corrections of kinematical origin.
A closer analysis shows that the $\cos \theta$-dependence of the (normalized)
angular distribution can be parametrized as \cite{Aubert:2006vb,Ali:1999mm,
  Bobeth:2001sq}
\begin{equation}
  \label{eq:babar}
  \frac{1}{\Gaml} \frac{d\Gaml }{d\!\cos\theta}
    = \frac{3}{4} (1 - \FHl) (1 - \cos^2\theta) 
    + \frac{1}{2} \FHl + \AFBl \cos\theta ,
\end{equation} 
with a flat term $\FHl/2$ and a linear term in $\cos \theta$, the
forward-backward asymmetry $\AFBl$. Both are small within the SM, and therefore
can signal the presence of NP. In particular they can be affected by Higgs and
tensor interactions.  Note that in the limit of vanishing lepton masses the SM
predicts the same rates for electrons and muons $\Game^{\rm SM}=\Gammu^{\rm SM}
+{\cal{O}}(m_\mu^2)$ if the same kinematical cuts are used \cite{Hiller:2003js}.

In this paper we analyze the angular distributions of $\BtoKll$ decays.  We
explicitly quantify the corrections to \refeq{eq:angular:SM} within the SM and
study model-independently the effects of $(\bar s b)( \bar l l)$ operators
induced by physics beyond the SM on \refeq{eq:babar}.  We use the framework of
QCD factorization (QCDF) valid in the low-$q^2$ region \cite{Beneke:2001at,
  Beneke:2004dp} and exploit the symmetries of QCD in the large recoil limit of
heavy-to-light transitions \cite{Charles:1998dr,Beneke:2000wa}.  Also, resonance
contributions from $\bar{B} \to K (c\bar{c}) \to K \bar{l}l$ can be controlled
for dilepton masses below the charm threshold.

The plan of the paper is as follows: After setting up the effective weak
Hamiltonian in \refsec{sec:eff:Ham} hadronic matrix elements are given in
\refsec{sec:hadr:ME}.  \refSec{sec:dec:dist} contains model-independent formulae
of the double differential and angular decay distributions.  We give numerical
predictions for the SM in \refsec{sec:SM:num} including a detailed discussion of
uncertainties. We also derive analytical expressions for $\Gaml$ and $\FHl$
obtained in the large recoil limit.  In \refsec{sec:NP} we work out the
sensitivity of the $\BtoKll$ angular distributions to NP in correlation with
other observables in $b \to s \bar l l$ decays.  We summarize in
\refsec{sec:conclusions}.  Technical details about form factors and form factor
symmetry relations in the low-$q^2$ region are given in \refapp{app:formf},
whereas details on the $\BtoKll$ hadronic matrix element in QCDF can be found in
\refapp{app:calT}.

%
%
%----------------------------------------------------
\section{The Effective Hamiltonian \label{sec:eff:Ham}}

The $\Delta B = 1$ effective Hamiltonian 
\begin{equation}
\label{eq:Heff}
  {\cal{H}}_{\rm eff}= 
  -\frac{4 G_F}{\sqrt{2}}  V_{tb}^{} V_{t s}^\ast \sum_i C_i(\mu)  \Op_i(\mu)
\end{equation}
is given in terms of dimension six operators $\Op_i$ and their respective Wilson
coefficients $C_i$. Both depend on the renormalization scale $\mu$, for which we
take a low energy scale $\mu_b$ of the order of the $b$-quark mass when
evaluating $B$-physics amplitudes. In \refeq{eq:Heff} the leading CKM elements
$V_{lm}$ are factored out. The sum over $i$ comprises the current-current
operators $i=1,2$, the QCD-penguin operators $i=3,4,5,6$, the photon and gluon
dipole operators $i=7,8$ and the semileptonic operators $i=9,10$. They are
defined as
\begin{align}
  \Op_7 & = 
    \frac{e}{(4 \pi)^2} \overline m_b [\bar{s} \sigma^{\mu\nu} P_R b] F_{\mu\nu}, &
  \Op_9 & = 
    \frac{e^2}{(4 \pi)^2} [\bar{s} \gamma_\mu P_L b][\bar{l} \gamma^\mu l], \nn
  \Op_8 & = 
    \frac{\gS}{(4 \pi)^2} \overline m_b [\bar{s} \sigma^{\mu\nu} P_R T^a b] G^a_{\mu\nu}, &
  \Op_{10} & = 
    \frac{e^2}{(4 \pi)^2} [\bar{s} \gamma_\mu P_L b][\bar{l} \gamma^\mu \gamma_5 l],  \label{eq:SMbasis}
\end{align}
where $P_{R/L} =(1 \pm \gamma_5)/2$ denote chiral projectors and $\overline
m_b(\mu)$ the $\MSbar$ $b$-quark mass at the scale $\mu$.  For the operators
$\Op_i$ with $i=1, \ldots ,6$ we use the definitions given in
\cite{Chetyrkin:1996vx}, also used by \cite{Beneke:2001at, Beneke:2004dp}. This
set of operators suffices to describe $b \to s \bar l l$ induced processes in
the SM, which are dominated by $C_{7},C_9$ and $C_{10}$, whereas $C_8$ enters at
higher order in the strong coupling.

Beyond the SM, NP might contribute in various ways. Assuming that NP manifests
itself at and above the electroweak scale, it can be model-independently
analyzed in the effective theory framework by allowing for NP contributions to
the Wilson coefficients of the SM operators and by additional operators not
present in the SM. To account also for the latter we include the most general $b
\to s$ (pseudo-) scalar and tensor operators with dileptons into our analysis:
\begin{align}
  \Op_S^l & = \frac{e^2}{(4 \pi)^2} [\bar{s} P_R b] [\bar{l} l], &
  \Op^{l  \prime}_S & = \frac{e^2}{(4\pi)^2} [\bar{s} P_L b] [\bar{l} l], 
\nonumber\\
  \Op_P^l & = \frac{e^2}{(4 \pi)^2} [\bar{s} P_R b] [\bar{l} \gamma_5 l], &
  \Op^{l  \prime}_P & = \frac{e^2}{(4 \pi)^2} [\bar{s} P_L b] [\bar{l} \gamma_5 l],    
\nonumber \\
  \Op_T^l   & = \frac{e^2}{(4 \pi)^2} [\bar{s} \sigma_{\mu\nu} b][\bar{l} \sigma^{\mu\nu} l], &
  \Op_{T5}^l & = \frac{e^2}{(4 \pi)^2} [\bar{s} \sigma_{\mu\nu} b][\bar{l} \sigma^{\mu\nu} \gamma_5 l],
  \label{eq:nonSM:op}
\end{align}
where we made the dependence on the lepton flavor explicit by the superscript
$l$.  Note that there are only two independent tensor operators
% with flavor structure $\bar s  b \bar l l$ 
in four dimensions.
At higher order also 4-quark operators with scalar, pseudoscalar and tensorial
structure contribute to rare radiative and semileptonic decays
\cite{Hiller:2003js,Borzumati:1999qt}.  Here we neglect these effects.

The additional NP operators \refeq{eq:nonSM:op} mix under QCD only with
themselves. Their 1-loop anomalous dimensions $\gamma_i = \frac{\alS}{4 \pi}
\gamma_i^{(0)}$
%($C_F=(N_C^2-1)/2 N_C$)
are
\begin{align} \label{eq:ad}
  \gamma_i^{(0)} &= - 6 \CF = - 8, &
  i & = S,S',P,P', &
 \nn
  \gamma_i^{(0)} &= 2 \CF = \frac{8}{3}, &
  i & = T,T5.
\end{align}
In our NP analyses all Wilson coefficients are taken at the low scale $\mu_b$.

%
%----------------------------------------------------
\section{The Hadronic Matrix Element at Large Recoil \label{sec:hadr:ME}}

A systematic treatment of the matrix element $\cM[\BtoKll] = \langle
l(p_-)\bar{l}(p_+) K(p_K) | {\cal{H}}_{\rm eff} | \bar{B}(p_B) \rangle$ is
available in the large recoil region.  We denote by $p_B,p_K,p_-$ and $p_+$ the
4-momenta of the $\bar B$-meson, kaon, lepton $l$ and antilepton $\bar l$,
respectively, and $M_B, M_K$ and $m_l$ are the corresponding masses.  At large
recoil the energy $E$ of the $K$-meson is large compared to the typical size of
hadronic binding energies $\LamConf \ll E$ and the dilepton invariant mass
squared $q^2 =(p_-+ p_+)^2$ is low, $q^2 \ll M_B^2$. Consequently, in this
region the virtual photon exchange between the hadronic part and the dilepton
pair and hard gluon scattering can be treated in an expansion in $1/E$ using
either QCDF or Soft Collinear Effective Theory (SCET) \cite{Bauer:2000yr}.
Furthermore, only one soft form factor $\xi_P(q^2)$ appears in the $\bar{B} \to
K$ heavy-to-light decay amplitude due to symmetry relations in the large energy
limit of QCD \cite{Charles:1998dr, Beneke:2000wa}.  Other nonperturbative
objects present are the light-cone distribution amplitudes (LCDAs) of the
$\bar{B}$- and $K$-mesons, leading to numerically smaller contributions. This
framework has been previously applied to $\BtoKastll$ decays using QCDF
\cite{Beneke:2001at, Beneke:2004dp} or SCET \cite{Ali:2006ew}.  In this work we
use the results from QCDF valid at low $q^2$ \cite{Beneke:2001at, Beneke:2004dp}
and include effects of finite lepton masses in $\BtoKll$ decays.

The $\BtoKll$ matrix element can be written as
\begin{align}
  \label{eq:matrix:el}
  \cM[\BtoKll] & = i \frac{\GF \alE}{\sqrt{2} \pi} V_{tb}^{} V_{ts}^{\ast}\,\, \xi_P(q^2) \,
     \Bigg( F_V\, p_B^{\mu}\, [\bar{l}\gamma_{\mu} l] + F_A\, p_B^{\mu}\,
     [\bar{l} \gamma_{\mu}\gamma_5 l] \\
  & \hspace{4.3cm} + (F_S + \cos \theta F_T) \, [\bar{l}l] 
      + (F_P + \cos \theta F_{T5}) \, [\bar{l}\gamma_5 l] \Bigg).
   \nonumber
\end{align}
Here, $\theta$ denotes the angle between the direction of motion of the
$\bar{B}$ and the negatively charged lepton $l$ in the dilepton center of mass
frame, following \cite{Bobeth:2001sq}. Note that this convention differs from
other works, e.g.,~\cite{Ali:1999mm}, where $\theta$ is defined with respect to
$\bar l$.  The functions $F_i \equiv F_i(q^2)$, $i=S,P,A,V,T,T5$ are given as
\begin{align}
  F_A & = C_{10}, &
  F_T & = \frac{2 \sqrt{\lambda}\, \beta_l}{M_B + M_K}\, \frac{f_T(q^2)}{f_+(q^2)}\, C_T^l\,, &
  F_{T5} & = \frac{2 \sqrt{\lambda}\, \beta_l}{M_B + M_K}\, \frac{f_T(q^2)}{f_+(q^2)}\, C_{T5}^l, 
 \nonumber
\end{align}
\vspace{3mm}
\begin{align}
  F_P & = \frac{1}{2} \frac{M_B^2-M_K^2}{m_b - m_s} \frac{f_0(q^2)}{f_+(q^2)} (C_P^l + C^{l \prime}_P)
        + m_l C_{10} \left[ \frac{M_B^2 - M_K^2}{q^2} \left(
         \frac{f_0(q^2)}{f_+(q^2)} - 1 \right) - 1\right], 
  \label{eq:Fis:def}
\end{align}
\vspace{3mm}
\begin{align}
  F_S & = \frac{1}{2} \frac{M_B^2 - M_K^2}{m_b - m_s} \frac{f_0(q^2)}{f_+(q^2)} (C_S^l + C^{l \prime}_S), &
  F_V & = C_9 + \frac{2 m_b}{M_B} \frac{{\cal T}_P(q^2)}{\xi_P(q^2)}
        + \frac{8 m_l}{M_B + M_K}\, \frac{f_T(q^2)}{f_+(q^2)}\, C_T^l,
 \nonumber
\end{align}
where
\begin{align}
\lambda &= M_B^4 + M_K^4 + q^4 - 2 (M_B^2 M_K^2 + M_B^2 q^2 + M_K^2 q^2), &
  \beta_l & = \sqrt{1 - 4 \frac{m_l^2}{q^2}},
\end{align}
and it is useful to note that $2 p_B \cdot(p_+ - p_-) = \sqrt{\lambda} \beta_l
\cos\theta$.  In the SM holds $F_S^{\rm SM} = F_T^{\rm SM} = F_{T5}^{\rm SM} =
0$.  Above, we have written the matrix element with the form factor $\xi_P(q^2)
=f_+(q^2)$ as an overall factor. It constitutes the main source of theoretical
uncertainties.  The form factor ratios $f_0/f_+$ and $f_T/f_+$ are constrained
by symmetry relations at large recoil \cite{Charles:1998dr,Beneke:2000wa}, which
are given in \refapp{app:formf} together with definitions of the form factors
and a discussion of their uncertainties.  The quantity ${\cal T}_P(q^2)$
appearing in the vector coupling to leptons, $F_V$, takes into account virtual
one-photon exchange between the hadrons and the lepton pair and hard scattering
contributions.  ${\cal T}_P(q^2)$ can be extracted from \cite{Beneke:2001at} and
is given in \refapp{app:calT}.  At lowest order
%in $\alpha_s$ and $1/m_b$ 
(denoted by the superscript $^{(0)}$) up to numerically small annihilation
contributions, it reads as
\begin{equation}
  {\cal T}_P^{(0)}(q^2) = \xi_P(q^2) \left[
     C_7^{\rm eff (0)} + \frac{M_B}{2 m_b} Y^{(0)}(q^2) \right].
\end{equation}
Hence, ${\cal T}_P(q^2)$ takes care of the contributions from the ${\cal{O}}_{1,
  \dots ,6}$ matrix elements $\propto Y(q^2)$ that are commonly included in an
effective coefficient of the operator ${\cal{O}}_9$ \cite{Buras:1994dj}. The
next-to leading $\alS$-corrections to ${\cal T}_P$ are known, see
\refapp{app:calT}, and taken into account in our analysis. Here we consider only
NP effects from the NP operators \refeq{eq:nonSM:op}, that is, their respective
coefficients as appearing in \refeq{eq:Fis:def} being non-zero, and ${\cal T}_P$
is SM-like.  The $b$-quark mass in ${\cal T}_P$ and $F_V$ is the potential
subtracted (PS) mass $m_b^{\rm PS}(\mu_f)$ at the factorization scale $\mu_f
\sim \sqrt{\LamConf m_b}$ and is denoted by $m_b$ throughout the paper. The
$b$-quark mass factors in $F_S$ and $F_P$ stem from the equations of motion, and
we take them in the PS scheme as well. In the evaluation of the function
$Y(q^2)$ we use the pole mass $m_b^{pole}$ \cite{Beneke:2001at}. The relation to
the PS mass is given as $m_{b}^{pole}=m_b^{\rm PS}(\mu_f)+ 4\alpha_s \mu_f/(3
\pi)$ \cite{Beneke:1998rk}.  The SM Wilson coefficients $C_9$ and $C_{10}$ are
taken in NNLL approximation \cite{Chetyrkin:1996vx,Bobeth:1999mk}. The remaining
SM Wilson coefficients $C_{1,\ldots ,6}$ and $C_{7,8}$ with their effective
counterparts $C_{7,8}^{\rm eff}$ enter only through ${\cal T}_P$.  For details
see \refapp{app:calT} and \cite{Beneke:2001at}.
Note that chirality flipped operators $\Op'_{7,9,10}$ can be readily included in
the matrix element of $\BtoKll$ decays by replacing $C_{9,10} \to C_{9,10}
+C^\prime_{9,10}$ in \refeq{eq:Fis:def} and $C_7 \to C_7 + C'_7$ in ${\cal
  T}_P$.

%
%----------------------------------------------------
\section{Decay Distributions of  $\BtoKll$ \label{sec:dec:dist}}

Based on the matrix element \refeq{eq:matrix:el} the double differential decay
rate with respect to $q^2$ and $\cos\theta$ with lepton flavor $l$ reads as
\begin{equation}
  \label{eq:double:dG}
  \frac{d^2\Gaml}{dq^2\, d\!\cos\theta} = 
    a_l(q^2) + b_l(q^2) \cos\theta + c_l(q^2) \cos^2\theta,
\end{equation}
where
\begin{align}
  \frac{a_l(q^2)}{\Gamma_0\, \sqrt{\lambda}\, \beta_l\, \xi_P^2} & = 
    q^2 \left( \beta^2_l |F_S|^2 + |F_P|^2 \right)
    + \frac{\lambda}{4}  (|F_A|^2 + |F_V|^2) 
  \nn
    & \hspace{2cm}
    + 2 m_l (M_B^2 - M_K^2 + q^2) Re(F_P F_A^\ast) + 4 m_l^2 M_B^2 |F_A|^2,
  \label{eq:a:of:s}
  \\[4mm] \label{eq:b:of:s}
  \frac{b_l(q^2)}{\Gamma_0\, \sqrt{\lambda}\, \beta_l\, \xi_P^2} & = 2\, \Big\{ 
    q^2 \left[ \beta_l^2 Re(F_S F_T^\ast) + Re(F_P F_{T5}^\ast) \right]
  \nn
    & \hspace{2cm} + m_l \left[ \sqrt{\lambda} \beta_l  Re(F_S F_V^\ast) 
       + (M_B^2 - M_K^2 + q^2) Re(F_{T5} F_A^\ast) \right] \Big\},
  \\[4mm]
  \frac{c_l(q^2)}{\Gamma_0\, \sqrt{\lambda}\, \beta_l\, \xi_P^2} & = 
    q^2 \left( \beta_l^2 |F_T|^2+ |F_{T5}|^2 \right)
    - \frac{\lambda}{4} \beta_l^2 (|F_A|^2 + |F_V|^2) 
    + 2 m_l \sqrt{\lambda} \beta_l Re(F_{T} F_V^\ast)
  \label{eq:c:of:s}
\end{align}
and 
\begin{equation}  \label{eq:norm}
  \Gamma_0 = \frac{\GF^2 \alE^2 |V_{tb}^{} V_{ts}^{\ast}|^2}{512 \pi^5 M_B^3}.
\end{equation}
These relations simplify considerably in the SM, where $b_l^{\rm SM}(q^2) = 0$
and in the limit $m_l \to 0$ further holds $a_l^{\rm SM}(q^2) = -c_l^{\rm
  SM}(q^2)$.

With \refeq{eq:double:dG} at hand the angular distribution
\begin{equation}
  \label{eq:dG:dtheta}
  \frac{d\Gaml}{d\cos\theta} = A_l + B_l \cos\theta + C_l \cos^2\theta
\end{equation}
is given in terms of the $q^2$-integrated coefficients
\begin{align} 
  A_l & = \int_{q^2_{\rm min}}^{q^2_{\rm max}} dq^2\, a_l(q^2), &
  B_l & = \int_{q^2_{\rm min}}^{q^2_{\rm max}} dq^2\, b_l(q^2), &
  C_l & = \int_{q^2_{\rm min}}^{q^2_{\rm max}} dq^2\, c_l(q^2).
\end{align} 
Their values depend on the cuts in $q^2$. We recall that while the boundaries of
the phase space allow for dilepton masses in the range $4 m_l^2 <q^2 \leq (M_B -
M_K)^2$, our calculation is valid only in the low-$q^2$ region. Note that for
very low dilepton masses there is sensitivity to light resonances. We therefore
restrict our analysis to $1 \GeV^2 \lesssim q^2 < 7 \GeV^2$.

The decay rate $\Gaml$ and the integrated and normalized forward-backward
asymmetry $\AFBl$ of the lepton pair can be expressed in terms of $A_l, B_l$ and
$C_l$
\begin{align}
  \Gaml & = 2 \left(A_l + \frac{1}{3} C_l \right), &
  \AFBl & = \frac{B_l}{\Gaml}.
  \label{eq:Gaml:AFBl}
\end{align}
We further introduce the observable
\begin{equation}
  \label{eq:FHl:def}
  \FHl \equiv \frac{2}{\Gaml} (A_l + C_l) 
  = {\dis \int_{q^2_{\rm min}}^{q^2_{\rm max}} dq^2\, \Big[a_l(q^2) + c_l(q^2)\Big]} \Bigg/
    {\dis \int_{q^2_{\rm min}}^{q^2_{\rm max}} dq^2\, \Big[a_l(q^2) + \frac{1}{3} c_l(q^2)\Big]}.
\end{equation}
With \refeq{eq:Gaml:AFBl} and \refeq{eq:FHl:def}, the angular distribution
\refeq{eq:dG:dtheta} is equivalent to \refeq{eq:babar} presented in the
Introduction. Since $\FHl$ is normalized to $\Gaml$, we expect reduced
uncertainties in the former compared to the latter due to cancellations between
numerator and denominator. As already anticipated after \refeq{eq:norm} within
the SM a cancellation takes place in \refeq{eq:FHl:def} between $a_l$ and $c_l$
such that $F_H^{l \, \rm SM}$ vanishes in the limit $m_l \to 0$. We discuss this
in detail in the next section. From here follows the approximate $\propto \sin^2
\theta$ angular dependence of $\BtoKll$ decays in the SM as in
\refeq{eq:angular:SM}.

We would like to comment on the possibility of corrections to \refeq{eq:babar}
or \refeq{eq:dG:dtheta} from higher powers of $\cos \theta$, that is, a
polynomial dependence in the angular distribution on $\cos^n\theta$ with $n >2$.
Higher angular momenta arise from higher $(>6)$ dimensional operators in the
weak Hamiltonian \refeq{eq:Heff} or from QED corrections. Hence, they are
suppressed by powers of external low energy momenta or masses over the
electroweak scale, and $\alpha_e/4 \pi$, respectively. We discuss such
corrections further at the end of \refsec{sec:SM:num} in the context of a
non-vanishing forward-backward asymmetry in $\BtoKll$ in the SM.

A further useful observable in $\BtoKll$ decays is $R_K$, the ratio of
$\BtoKmumu$ to $\BtoKee$ decay rates with the same $q^2$ cuts
\cite{Hiller:2003js}
\begin{align}
  \label{eq:RK:def}
  R_K & \equiv 
   \frac{\Gammu}{\Game}
  =
  {\dis \int_{q^2_{\rm min}}^{q^2_{\rm max}} dq^2\, \frac{d\Gammu}{dq^2}}\Bigg/
  {\dis \int_{q^2_{\rm min}}^{q^2_{\rm max}} dq^2\, \frac{d\Game}{dq^2}}
  = \frac{\Gammu \FHmu - 4/3\, C_\mu}{\Game},
\end{align}
which probes lepton flavor dependent effects in and beyond the SM. We find that
$\FHl$ and $R_K$ are model-independently related
\begin{align}
\label{eq:RKvsFH}
  R_K \cdot ( 1 - \FHmu - \Delta) & = 1, & \mbox{where}~~~~~
  \Delta = & \frac{4}{3} \frac{C_e - C_\mu}{\Gammu} - \frac{\FHe}{R_K}.
\end{align}
The expression for $\Delta$ simplifies in models where chiral couplings to
electrons can be neglected as, for example, in the SM with $m_e=0$.  Then $\FHe
= 0$ and $\Game = -4/3 C_e$ and in the SM $\Delta^{\rm SM} \propto m_\mu^2$. We
carefully examine SM predictions for $\Gaml, \FHmu$ and $R_K$ in
\refsec{sec:SM:num} and work out the NP potential of $R_K$, $\FHl$ and $\AFBl$
in \refsec{sec:NP}. Corresponding values for $\Delta$ can be obtained by means
of \refeq{eq:RKvsFH}.

%
%
%----------------------------------------------------
\section{Standard Model Predictions \label{sec:SM:num}}

In this section we analyze $\BtoKll$ decays within the SM. We give predictions
for the observables $\FHl$, $R_K$ and $\Gaml$ and the corresponding branching
ratios $\BRl \equiv {\cal{B}}(\BtoKll)$ for low dilepton mass.  Higher order SM
contributions to the forward-backward asymmetry $\AFBl$ are briefly discussed.

%
% Width
%

We start with a general analysis of lepton flavor dependence in the $\BtoKll$
decay rate $\Gaml$. In the SM, such effects are of purely kinematical origin and
often negligible \cite{Hiller:2003js}. At large recoil, the suppression of the
lepton mass induced terms can be quantified analytically using form factor
symmetry relations \refeq{eq:ff:sym:rel}.  For low $q^2$, $\Gaml$ can then be
written as
\begin{align}
  \label{eq:Gaml:SM}
  \Gaml^{\rm SM} & = \frac{\Gamma_0}{3} \int_{q^2_{\rm min}}^{q^2_{\rm max}} dq^2\, 
  \xi_P^2(q^2) \sqrt{\lambda}^3 (|F_A|^2 + |F_V|^2) 
\\
  &\hspace{2cm} \times\Bigg\{ 1 + \order{\frac{m_l^4}{q^4}} + \frac{m_l^2}{ M_B^2} 
   \times \order{ \alS, \frac{q^2}{M_B^2} \sqrt{\frac{\LamConf}{E}}} \Bigg\},
 \nonumber
\end{align}
where the $m_l^4$ correction has been obtained from explicit expansion of the
coefficients $a_l$ \refeq{eq:a:of:s} and $c_l$ \refeq{eq:c:of:s} in $m_l$. (A
useful relation is given in \refeq{eq:low:q2:rel}.) Due to a cancellation with
the kinematical function $\beta_l$ there are no terms of order $m_l^2$ up to
symmetry breaking corrections, which are estimated in the second correction term
in \refeq{eq:Gaml:SM}. Form factor relations are broken in general by
$\alS$-corrections and power corrections in $\LamConf/E$, as discussed in more
detail in \refapp{app:formf}.  As can be seen, these receive here further strong
suppression from $m_l^2/M_B^2$. Note that consistent with the $\LamConf/E$
expansion we neglected terms of order $M_K^2/M_B^2$ and we approximated in the
symmetry breaking correction contribution in \refeq{eq:Gaml:SM} $\lambda \approx
M_B^4$, thereby dropping terms suppressed by $q^2/M_B^2$.

We conclude from \refeq{eq:Gaml:SM} that lepton mass effects in the SM $\BtoKll$
decay rate at low $q^2$ are of order $m_\mu^4/q^4 \sim 10^{-4}$ for muons and
even further down by $m_e^2/m_\mu^2 \simeq 2 \cdot 10^{-5}$ for electrons, hence
negligible in agreement with earlier numerical findings covering the whole
dilepton mass region \cite{Hiller:2003js}.  To leading order in $m_l$, the decay
rate depends then only on $|F_V|$ and $|F_A|$.  The function $F_A$ equals the
Wilson coefficient $C_{10}^{\rm SM} \sim -4$ with tiny dependence on the low
scale $\mu_b$.  $F_V$ is a sum of $C_9^{\rm SM} \sim +4$ and a term containing
${\cal T}_P$.  The latter is subject to unknown higher order power corrections.
However, the typical order of magnitude of $|{\cal T}_P(q^2)| \sim 0.1$ implies
that these corrections constitute a rather small contribution to $F_V$ and the
corresponding uncertainties are very small compared to the dominating one from
the overall form factor $\xi_P$.

\begin{figure}
\begin{center}
  \epsfig{figure=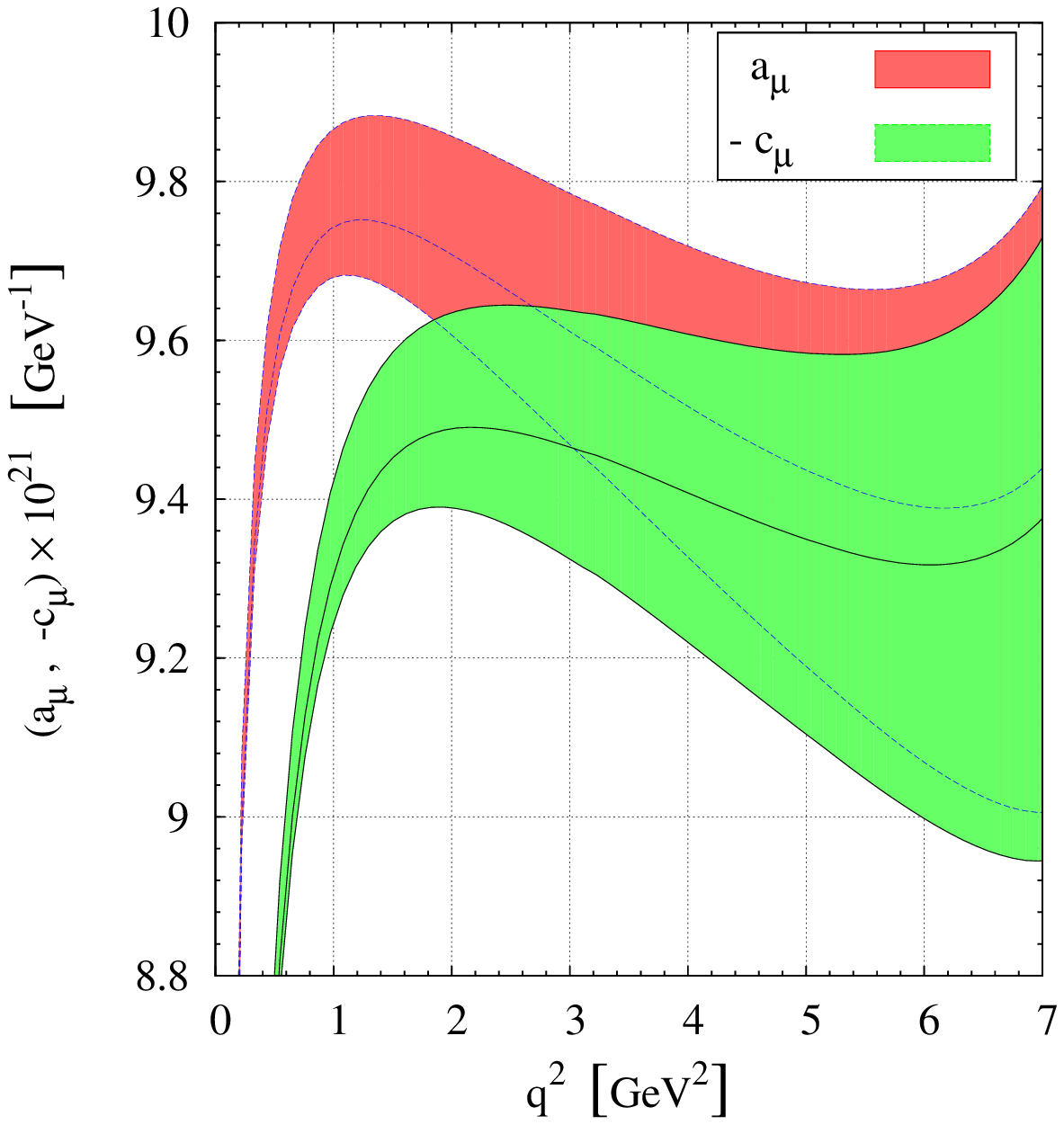, height=7.5cm, angle=-0} 
  \epsfig{figure=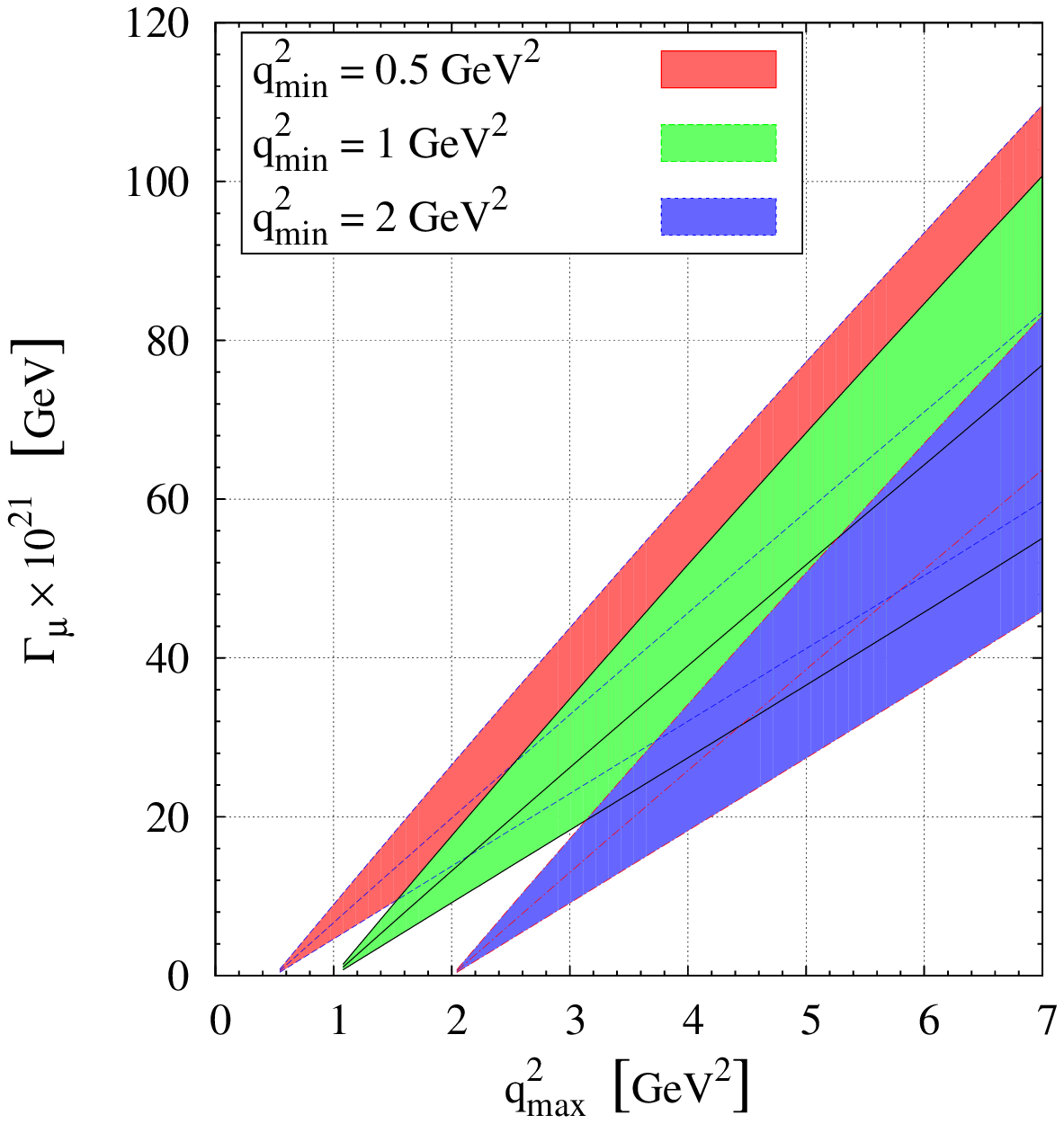,  height=7.5cm, angle=-0}
\end{center}
\caption{ \label{fig:ac:Gammu} In the left-hand plot $a_l(q^2)$ and $-c_l(q^2)$
  defined in \refeq{eq:double:dG} are shown for $l = \mu$ in the SM as a function
  of $q^2$ for the renormalization scale $\mu_b$ between $m_b/2$ and $2 m_b$. In
  the right-hand plot the SM $\BtoKmumu$ decay rate is given for three different
  cuts $q^2_{\rm min} = \{0.5, 1, 2\} \GeV^2$ as a function of $q^2_{\rm max}$.
  Here the bands take into account uncertainties from the form factor $\xi_P$,
  $\mu_b$ and $V_{ts}$. }
\end{figure}

The numerical analysis of $\Gaml$ confirms the discussed qualitative features.
The main uncertainties are due to the form factor $\xi_P$, the CKM matrix
element $V_{ts}$ and the renormalization scale $\mu_b$.  For the form factor
$\xi_P(q^2)$ we use the findings from Light Cone Sum Rules (LCSR)
\cite{Ball:2004ye}. At low dilepton mass, the form factor has an uncertainty
between $(12- 16) \%$, with smaller uncertainty for larger $q^2$, for details
see \refapp{app:formf}.  Our numerical input is given in \reftab{tab:num:input}.
We find that the $\mu_b$-dependence of the decay rate is rather small, about a
few percent, as can be seen from \reffig{fig:ac:Gammu} (left-hand plot). Here
the coefficients $a_\mu(q^2)$ and, to enable easier comparison, $-c_\mu(q^2)$
are shown for $\mu_b$ between $m_b/2$ and $2 m_b$. The small uncertainty due to
$\mu_b$ is not unexpected because of the inclusion of NNLL corrections to the
matrix elements of the current-current operators \cite{Asatrian:2001de} in
${\cal T}_P$, which cancels the $\mu_b$-dependence of $C_9^{\rm SM}$.  In the
right-hand plot of \reffig{fig:ac:Gammu} we show $\Gammu$ for three lower cuts
$q^2_{\rm min} = \{0.5, 1, 2 \} \GeV^2$ as a function of the upper boundary
$q^2_{\rm max}$. The combined uncertainty from $\xi_P(q^2)$, $\mu_b$ and
$V_{ts}$ can be as large as $32\%$.  Further subleading sources are the 
lifetime
with $0.7 \%$ uncertainty and $\alpha_e(\mu)$, which enters quadratically and
brings in about $6 \%$ uncertainty to the $\BtoKll$ decay rates.  
The latter can be reduced by including the higher order electroweak
corrections from \cite{Bobeth:2003at,Huber:2005ig} 
to the renormalization group evolution
which should capture the leading effect. For a complete higher order
electroweak analysis the QED-corrections to the $\BtoKll$ matrix element
should be calculated. In
the corresponding calculation for inclusive $\BXsll$ decays collinear
logarithms of order $\alpha_e/(4 \pi)  \cdot \log(m_l/m_b)$ 
arise for low dilepton
mass cuts \cite{Huber:2005ig}. The resulting splitting between electron and
muon final states, however, diminishes after experimental cuts which
separate electrons from energetic collinear photons.
How much this matters for $\Game$ and $\Gammu$ and $R_K$ cannot be answered
until these QED-corrections are calculated. The uncertainties in $\Gaml$ from 
the charm, bottom and top mass are $2 \%, 0.4 \%$ and $2 \%$, respectively.

\begin{table}
%\TABLE[ht]{
\begin{center}
\begin{tabular}{||l|l||}
\hline \hline
  $\alS(m_Z) = 0.1176 \pm 0.0020$ \hfill \cite{Yao:2006px}
&
  $f_K = (159.8 \pm 1.4 \pm 0.44) \MeV$  \hfill \cite{Yao:2006px} 
\\
  $\alE(m_b) = 1/133 $ 
&
  $f_{B_{u,d}} = (200 \pm 30) \MeV$ \hfill 
\\
 $|V_{ts}| = 0.0409 \pm 0.0021 $ \hfill \cite{CKMfitter:homepage} 
& 
  $f_{B_{s}} = (240 \pm 30) \MeV$ \hfill \cite{Onogi:2006km}
\\
 $|V_{cb}| = 0.0416 \pm 0.0007 $ \hfill \cite{CKMfitter:homepage} 
&
  $a_1^K(1 \GeV) = 0.06 \pm 0.03$ \hfill \cite{Ball:2006wn} 
\\
  $m_W = 80.403  \GeV$ \hfill \cite{Yao:2006px}
&
  $a_2^K(1 \GeV) = 0.25 \pm 0.15$ \hfill \cite{Ball:2006wn} 
\\
  $m_t^{pole} = (170.9 \pm 1.8) \GeV$ \hfill \cite{:2007bx}
& 
$a_4^K(1 \GeV) = -0.015 \pm 0.1$ \hfill \cite{Ball:2004ye} 
\\
  $m_b=(4.6 \pm 0.1) \GeV$ \hfill \cite{Beneke:2001at}
&
  $\lambda_{B,+}(1.5 \GeV) = (0.458 \pm 0.115) \GeV$ \hfill 
\cite{Beneke:2004dp, Braun:2003wx} 
\\
 $m_c^{pole}=(1.4 \pm 0.2) \GeV$
& $\xi_P(0)=0.327 \pm 0.053$   \hfill \cite{Ball:2006wn,Ball:2004ye} 
\\
 $\BR(\BXclv)=(10.57 \pm 0.15) \%$  \hfill\cite{Yao:2006px}
& $\tau_{B^\pm} = (1.638 \pm 0.011) \, {\rm ps}$ \hfill \cite{Yao:2006px}
\\
 & $\tau_{B^0} = (1.530 \pm 0.009) \,  {\rm ps} $ \hfill\cite{Yao:2006px} 
\\
& $\tau_{B_s} = (1.425 \pm 0.041) \,  {\rm ps} $ \hfill\cite{Yao:2006px}
\\
\hline \hline
\end{tabular}
\end{center}
\caption{ \label{tab:num:input} The numerical input used in our analysis.
  We denote by $m_b$ the PS mass at the factorization scale $\mu_f=2 \GeV$.
  We neglect the strange quark mass throughout this work. }
\end{table}

In \reftab{tab:SM:res} predictions for the SM branching ratios of $\BmtoKmmumu$
and $\B0toK0mumu$ decays are given including the uncertainties from
$\xi_P(q^2)$, $V_{ts}$ and $\mu_b$ added in quadrature.  The relative errors due
to $\xi_P$ and $\mu_b$ are given also separately.  Lepton mass effects are
negligible in $\Gaml$ and $\BRl$, and the decay rates and branching ratios with
electrons agree within uncertainties with the corresponding ones with muons.
The splitting of $(9.5-9.7) \%$ between the branching ratios of the $B^-$ and
$\bar B^0$ mesons is dominated by the lifetime difference, but there is also a
small isospin breaking contribution from spectator effects residing in ${\cal
  T}_P$.

\TABLE[ht]{
\begin{tabular}{||c|c c c|c c c||}
\hline \hline
\multirow{2}{*}{} & \multicolumn{3}{c|}{$\BmtoKmll$} & \multicolumn{3}{c||}{$\BneutralKll$}
\\
\cline{2-7}
          & SM value & $\quad\xi_P [\%]\quad$ & $\quad\mu_b [\%]\quad$
          & SM value & $\quad\xi_P [\%]\quad$ & $\quad\mu_b [\%]\quad$
\\
\hline \hline
\multirow{2}{*}{$\BRmu$}
& ${1.60}_{ - 0.46}^{ + 0.51}$ & ${}_{ - 27.0}^{ + 29.9}$ & ${}_{ - 1.8}^{ +
2.0}$
& ${1.46}_{ - 0.43}^{ + 0.47}$ & ${}_{ - 27.4}^{ + 30.4}$ & ${}_{ - 2.0}^{ +
2.1}$
\\[0.5ex]
& ${1.27}_{ - 0.36}^{ + 0.40}$ & ${}_{ - 26.6}^{ + 29.4}$ & ${}_{ - 2.1}^{ +
2.2}$
& ${1.16}_{ - 0.33}^{ + 0.37}$ & ${}_{ - 27.0}^{ + 29.8}$ & ${}_{ - 2.2}^{ +
2.3}$
\\[0.5ex]
\multirow{2}{*}{$[10^{-7}]$}
& ${1.91}_{ - 0.54}^{ + 0.59}$ & ${}_{ - 26.6}^{ + 29.2}$ & ${}_{ - 2.2}^{ +
2.2}$
& ${1.74}_{ - 0.50}^{ + 0.55}$ & ${}_{ - 26.8}^{ + 29.6}$ & ${}_{ - 2.3}^{ +
2.3}$
\\[0.5ex]
& ${1.59}_{ - 0.44}^{ + 0.48}$ & ${}_{ - 26.0}^{ + 28.7}$ & ${}_{ - 2.4}^{ +
2.4}$
& ${1.45}_{ - 0.41}^{ + 0.45}$ & ${}_{ - 26.3}^{ + 29.0}$ & ${}_{ - 2.6}^{ +
2.5}$
\\[0.5ex]
\hline
\multirow{4}{*}{$\FHmu$}
& ${0.0244}_{ - 0.0003}^{ + 0.0003}$ & ${}_{ - 1.0}^{ + 0.8}$ & ${}_{ - 0.5}^{ +
0.7}$
& ${0.0243}_{ - 0.0003}^{ + 0.0003}$ & ${}_{ - 1.1}^{ + 0.9}$ & ${}_{ - 0.4}^{ +
0.7}$
\\[0.5ex]
& ${0.0188}_{ - 0.0001}^{ + 0.0002}$ & ${}_{ - 0.5}^{ + 0.4}$ & ${}_{ - 0.4}^{ +
0.7}$
& ${0.0187}_{ - 0.0001}^{ + 0.0002}$ & ${}_{ - 0.5}^{ + 0.5}$ & ${}_{ - 0.4}^{ +
0.7}$
\\[0.5ex]
& ${0.0221}_{ - 0.0003}^{ + 0.0003}$ & ${}_{ - 1.4}^{ + 1.2}$ & ${}_{ - 0.6}^{ +
0.9}$
& ${0.0221}_{ - 0.0004}^{ + 0.0003}$ & ${}_{ - 1.5}^{ + 1.2}$ & ${}_{ - 0.6}^{ +
0.9}$
\\[0.5ex]
& ${0.0172}_{ - 0.0002}^{ + 0.0002}$ & ${}_{ - 0.8}^{ + 0.7}$ & ${}_{ - 0.6}^{ +
0.9}$
& ${0.0172}_{ - 0.0002}^{ + 0.0002}$ & ${}_{ - 0.8}^{ + 0.7}$ & ${}_{ - 0.6}^{ +
0.9}$
\\[0.5ex]
\hline
\multirow{4}{*}{$R_K$}
& ${1.00030}_{ - 0.00007}^{ + 0.00010}$ & ${}_{ - 0.003}^{ + 0.004}$ & ${}_{ -
0.006}^{ + 0.010}$
& ${1.00031}_{ - 0.00007}^{ + 0.00010}$ & ${}_{ - 0.003}^{ + 0.004}$ & ${}_{ -
0.006}^{ + 0.010}$
\\[0.5ex]
& ${1.00037}_{ - 0.00007}^{ + 0.00010}$ & ${}_{ - 0.003}^{ + 0.004}$ & ${}_{ -
0.006}^{ + 0.010}$
& ${1.00038}_{ - 0.00007}^{ + 0.00011}$ & ${}_{ - 0.003}^{ + 0.004}$ & ${}_{ -
0.006}^{ + 0.010}$
\\[0.5ex]
& ${1.00032}_{ - 0.00007}^{ + 0.00010}$ & ${}_{ - 0.003}^{ + 0.004}$ & ${}_{ -
0.006}^{ + 0.010}$
& ${1.00033}_{ - 0.00007}^{ + 0.00011}$ & ${}_{ - 0.003}^{ + 0.004}$ & ${}_{ -
0.006}^{ + 0.010}$
\\[0.5ex]
& ${1.00039}_{ - 0.00007}^{ + 0.00011}$ & ${}_{ - 0.003}^{ + 0.004}$ & ${}_{ -
0.006}^{ + 0.010}$
& ${1.00040}_{ - 0.00007}^{ + 0.00011}$ & ${}_{ - 0.003}^{ + 0.004}$ & ${}_{ -
0.007}^{ + 0.010}$  \\[0.5ex]
\hline \hline
\end{tabular}
\caption{ \label{tab:SM:res} SM predictions for $\BRmu$ (in units of $10^{-7}$),
  $\FHmu$ and $R_K$ for charged and neutral $B$-meson decays and different $q^2$
  cuts $(q^2_{\rm min}, q^2_{\rm max}) = (1, 6), (2, 6), (1, 7), (2, 7) \GeV^2$
  (from top to bottom).  The uncertainties from the form factor $\xi_P(q^2)$ and
  the renormalization scale $\mu_b$ varied between $m_b/2$ and $2 m_b$ are also
  given separately in percent of the central value.  The corresponding branching
  ratios with electrons, $\BRe$, agree within uncertainties with the ones with
  muons, $\BRmu$.  For details see text.}}
%\end{table}

%
%  R_K
%

\begin{figure}
\begin{center}
  \epsfig{figure=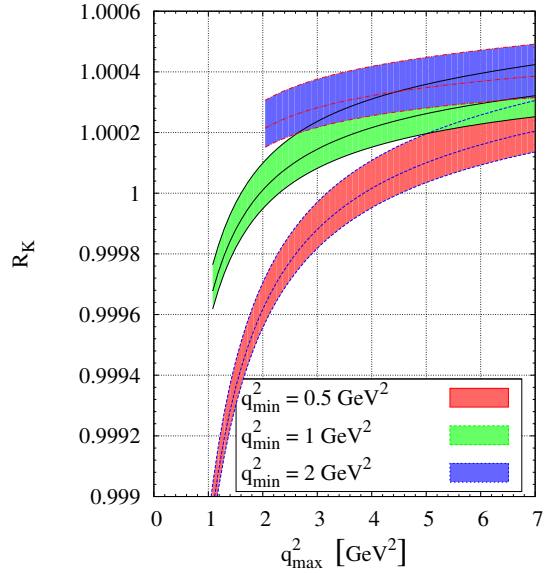,      height=8cm, angle=-0} 
\end{center}
\caption{ \label{fig:RK:SM} The ratio $R_K$ in the SM for different cuts
  $q^2_{\rm min} = \{0.5, 1, 2\} \GeV^2$ as a function of $q^2_{\rm max}$. The
  uncertainties from the scale $\mu_b$ and the form factor are added in
  quadrature. }
\end{figure}

In view of the insensitivity of $\Gaml^{\rm SM}$ to lepton mass effects for
$l=e, \mu$ and with regard to its large form factor uncertainty it was proposed
in \cite{Hiller:2003js} to investigate the ratio $\Gammu/\Game$, i.e., $R_K$
\refeq{eq:RK:def}. Our numerical analysis confirms a cancellation of the
hadronic uncertainties in $R_K$ also for low dilepton mass as can be seen in
\reffig{fig:RK:SM}. Here we show $R_K$ for different cuts $q^2_{\rm min} = \{0.5,
1, 2\} \GeV^2$ versus $q^2_{\rm max}$. The combined uncertainty due to the form
factor $\xi_P(q^2)$ and the renormalization scale $\mu_b$ is given by the bands
and is tiny. This can be seen also from \reftab{tab:SM:res}. The $\mu_b$ and
$\xi_P(q^2)$ induced uncertainties in $R_K$ are of comparable size, of the order
$\lesssim 10^{-4}$.  The deviation of $R_K^{\rm SM}$ from 1 is mainly due to the
inclusion of effects of $\order{m_\mu^4/q^4} \sim 10^{-4}$ given in
\refeq{eq:Gaml:SM}.  Any measured deviation of $R_K$ from 1 thus will signal NP
which does not contribute equally to $\Gammu$ and $\Game$ as, for example, in
the presence of non-universal lepton couplings.

%
%  F_H^l
%

\begin{figure}
\begin{center}
  \epsfig{figure=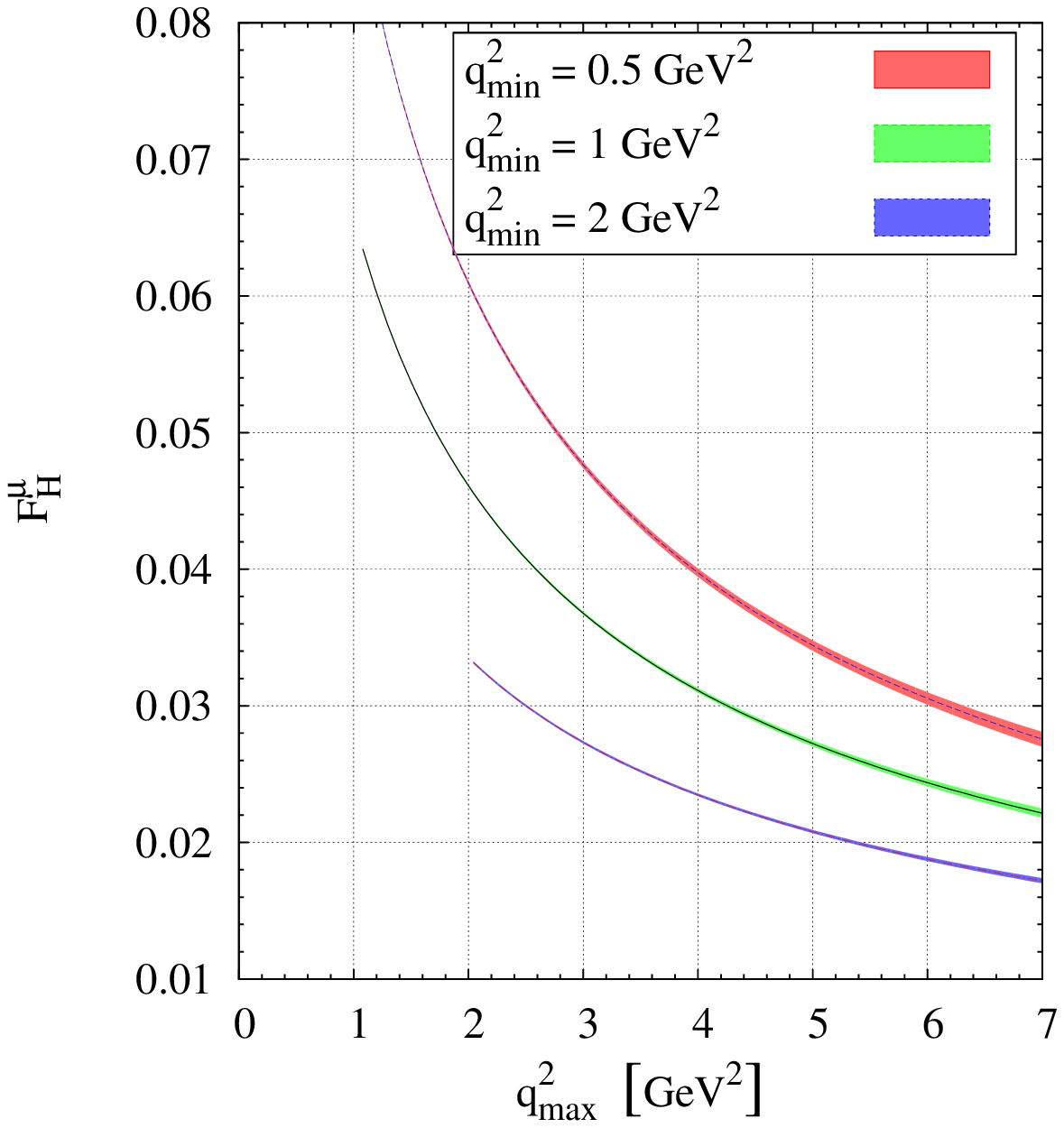,      height=7.5cm, angle=-0} 
  \epsfig{figure=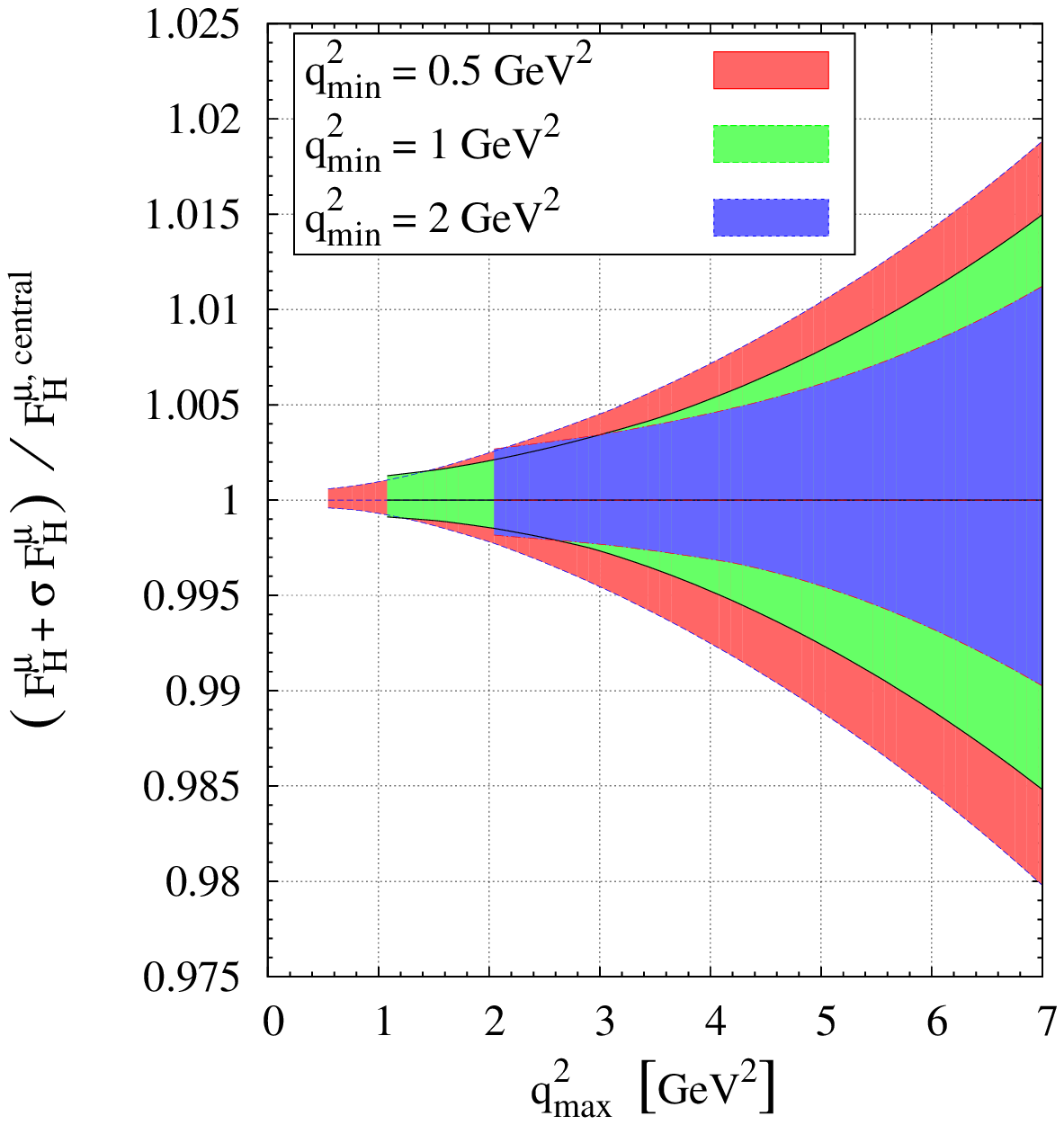, height=7.5cm, angle=-0}
\end{center}
\caption{ \label{fig:FHmu} The observable $\FHmu$ in the SM depending on
  $q^2_{\rm max}$ for three cuts $q^2_{\rm min} = \{0.5, 1, 2\} \GeV^2$ (left-hand
  plot) and normalized to the central value (right-hand plot).  The bands
  include combined uncertainties from $\mu_b$ and the form factor $\xi_P(q^2)$.  }
\end{figure}

Similar to $R_K$ also the angular observable $\FHl$ \refeq{eq:FHl:def} is a
ratio, where the overall factor $\Gamma_0$ \refeq{eq:norm} drops out and
uncertainties can cancel.  With the aid of the form factor symmetry relations
\refeq{eq:ff:sym:rel} and \refeq{eq:low:q2:rel} we obtain a simple expression
for $\FHl$ in the SM at low $q^2$:
\begin{align}
  \label{eq:FHl:SM}
  F_H^{l \, \rm SM} & =2  m_l^2 \frac{\Gamma_0}{\Gaml^{\rm SM}} \int_{q^2_{\rm min}}^{q^2_{\rm max}} \frac{dq^2}{q^2} \,
       \xi_P^2(q^2) \sqrt{\lambda}^3 \beta_l (|F_A|^2 + |F_V|^2) 
  \\  & \hspace{3cm} \times
    \Bigg\{ 1 + \frac{q^2}{M_B^2} \times 
    \order{ \alS, \frac{q^2}{M_B^2} \sqrt{\frac{\LamConf}{E}}} \Bigg\},
  \nonumber
\end{align}
where the denominator $\Gaml^{\rm SM}$ is given in \refeq{eq:Gaml:SM}.  {}From
the lepton mass suppression of $a_l+c_l$ in the numerator of \refeq{eq:FHl:def}
follows $F_H^{l \, \rm SM} \propto m_l^2$, and $F_H^{e \, \rm SM}/F_H^{\mu \,
  \rm SM} \propto m_e^2/m_\mu^2$ such that $F_H^{e \, \rm SM}$ is negligible.
The cancellation between $a_\mu(q^2)$ and $c_\mu(q^2)$ is also visible from
\reffig{fig:ac:Gammu} (left-hand plot).  Note that the leading term of the
integrand in the numerator of \refeq{eq:FHl:SM} is the same as the one in the
denominator \refeq{eq:Gaml:SM} except for an additional factor of $\beta_l/q^2
\simeq 1/q^2$. We therefore expect large cancellations of uncertainties in the
ratio for low $q^2$. This concerns the ones from the form factor, the
renormalization scale, $V_{ts}$ and unknown subleading $1/E$ corrections in
${\cal T}_P$.

As expected the SM values of $\FHmu$ are rather small, i.e., at the percent
level, with the exact value depending on cuts.  This can be seen from
\reffig{fig:FHmu}, where $\FHmu$ is shown for $q^2_{\rm min}= \{0.5,\, 1,\, 2\}
\GeV^2$ versus the upper integration boundary $q^2_{\rm max}$.  For $q^2_{\rm
  min} = \{1,\, 2\} \GeV^2$, $\FHmu$ ranges between $0.015 - 0.05$ depending on
the values of $q^2_{\rm min}$ and $q^2_{\rm max}$.  $\FHmu$ becomes larger for
smaller dilepton mass intervals and also for lower values of the lower cut
$q^2_{\rm min}$.  SM values of $\FHmu$ are given for some low-$q^2$ cuts in
\reftab{tab:SM:res}. Within uncertainties, the predictions for $\BmtoKmll$ and
$\BneutralKll$ decays are the same.

Indeed our numerical analysis of $\FHmu$ exhibits strong cancellations of
uncertainties. The form factor $\xi_P$ and $\mu_b$ induce uncertainties of
comparable sizes of order one percent, see \reftab{tab:SM:res}.  The combined
uncertainty from $\xi_P(q^2)$ and $\mu_b$ is indicated by the small bands in
\reffig{fig:FHmu} and result in an $\lesssim 2\%$ uncertainty, see also
\reftab{tab:SM:res}.  Power counting suggests an additional uncertainty from
form factor symmetry breaking of order $q^4/M_B^4 \sqrt{\LamConf/E} \sim 3 \%$
in $\FHmu$.  We also allow for subleading power corrections to the hard
scattering contributions at the order $q^2/M_B^2 \alpha_s \sqrt{\LamConf/E} \sim
3 \%$, see \refapp{app:formf}.  Taking all this into account, $\FHmu$ can be
predicted with an accuracy of $\sim {\cal{O}}(6 \%)$ in the SM, which is a high
precision for an observable in exclusive $B$-decays. Due to its huge suppression
from $m_e^2$, $\FHe$ is a null test of the SM.  Comparing our SM predictions for
$R_K$ and $\FHmu$, the former is known even more precisely due to the
cancellation of the ${\cal{O}}(m_l^2)$-terms at leading order in $\alpha_s$ and
$1/E$ and the stronger suppression of the symmetry relation breaking corrections
in $\Gaml$ \refeq{eq:Gaml:SM} compared to the ones in \refeq{eq:FHl:SM}.  In
order for the relation \refeq{eq:RKvsFH} to hold, $\Delta^{\rm SM}$ must be
equal to $-F_H^{\mu \, \rm SM}$ at the level of $10^{-4}$.

As already mentioned in \refsec{sec:dec:dist}, operators in the effective theory
of dimension higher than six or QED corrections induce additional contributions
to the $\BtoKll$ decay amplitude, which can modify the angular distributions.
As for the higher dimensional operators, in the SM they are, for example,
generated at one-loop by the Higgs penguin and the box with one charged pseudo
Goldstone- and one $W$-boson \cite{Krawczyk:1989qp}.  Contributions to scalar
and pseudoscalar operators arise then at the order $C_{S,P}^{l \, \rm SM} \sim
m_l m_b/m_W^2$.  Plugging this into \refeq{eq:Fis:def}, \refeq{eq:b:of:s} and
\refeq{eq:Gaml:AFBl}, a non-zero forward-backward asymmetry $A_{\rm FB}^{l \,
  \rm SM} \sim m_l^2/m_W^2$ is induced, which is too small to be experimentally
probed.  The corresponding SM tensor contributions have not been calculated, but
they are subject to a similar ${\cal{O}}( m_l m_b/m_W^2)$ suppression, and
negligible as the scalar ones in the $\BtoKll$ observables.

Higher order $\alpha_e$-corrections to exclusive $\BtoKll$ decays have not been
considered so far. Besides reducing the uncertainty from the overall
$\alpha_e(\mu)$ in the decay amplitude, radiative corrections can distort the
decay distributions at the level of $\alE/(4 \pi)$.  The generation of an
interesting $\cos \theta$-behavior from QED has been demonstrated for $K \to \pi
\bar e e$ decays. Radiative corrections via $K \to \pi \gamma \gamma$ enter the
$\bar e \gamma_\mu e$-form factor in the matrix element,
e.g.,~\cite{Donoghue:1994yt}, which can be parametrized in our notation as $F_V
\to F_V + \alE/(4 \pi) \cos \theta \tilde F_V$, see \refeq{eq:matrix:el}. Note
that $F_V, \tilde F_V$ are functions of $q^2$ only. The extra power of $\cos
\theta$ implies not only a non-zero $A_{\rm FB}^{l \, \rm SM} \sim \alE/(4 \pi)
\tilde F_V/C_9^{\rm SM} $, but also a $\cos^{3} \theta$-term of order $\alE/(4
\pi) \tilde F_V C_9^{\rm SM}$ and a suppressed $\cos^{4} \theta$-term of order
$(\alE/(4 \pi) \tilde F_V)^2$ in the angular distributions. Unless the unknown
correction factor $\tilde F_V$ is significantly enhanced ($ \gg 1$), it is
unlikely that $\alpha_e$-corrections have observable consequences in $\BtoKll$
decays.

%
%
%----------------------------------------------------
\section{Beyond the Standard Model \label{sec:NP}}

In the first part of this section we perform a model-independent analysis of
$\BtoKll$ decays for $l=e$ and $l=\mu$.  The size of the deviations from the SM
in the observables $\FHl$, $R_K$ and $\AFBl$ due to the NP operators
\refeq{eq:nonSM:op} is estimated. We show this for four benchmark scenarios in
\refsec{sec:scen1} to \refsec{sec:scen4}.  The second part of this section,
\refsec{sec:tensors}, contains a brief discussion of models with (pseudo-)
scalar and tensor operators.  All NP Wilson coefficients are assumed to be real
and are understood to be at the low scale $\mu_b$, i.e., here $C_i^l
= C_i^l(\mu_b)$.  Leading logarithmic renormalization group evolution to the
electroweak scale can be done with the anomalous dimensions given in
\refeq{eq:ad}.

We start with some general considerations about the dependence of the $\BtoKll$
observables on the NP Wilson coefficients.  Up to corrections of order $m_l^3$
we find for the branching ratio
\begin{align}
  \BRl = & \left[ \frac{\tau_{B^\pm}}{1.64 \rm ps} \right]
    \Bigg[1.91 + 0.02\,(C_S^{l 2} + C_P^{l 2}) + 0.06\, (C_T^{l 2} + 
C_{T5}^{l 2}) 
    + \frac{m_l}{\GeV} \Big( \frac{C_T^l}{0.99} - \frac{C_P^l}{2.92} \Big)  
\nonumber \\ 
  & + \frac{m_l^2}{\GeV^2} \Big( \frac{C_T^{l 2}}{3.28^2} - \frac{C_{T5}^{l 2}}{ 3.28^2}
     - \frac{C_P^{l 2}}{10.36^2} - \frac{C_S^{l 2}}{5.98^2} \Big) 
    + \order{m_l^3} \Bigg] \cdot 10^{-7},
  \label{eq:BRl:NP}
\end{align}
the numerator of $\FHl$ \refeq{eq:FHl:def}
\begin{align}
  \label{eq:FHlnum:NP}
  2\, \tau_{B^\pm}\, (A_l + C_l) = & \left[ \frac{\tau_{B^\pm}}{1.64 \rm ps} \right] 
  \Bigg[ \frac{m_l^2}{(0.51 \GeV)^2}+ 0.02\, (C_S^{l 2} + C_P^{l 2}) + 0.19\,(C_T^{l 2} + C_{T5}^{l 2} )
\\
   + \frac{m_l}{\GeV} \Big( \frac{C_T^l}{0.99} - \frac{C_P^l}{2.92} \Big)
  & + \frac{m_l^2}{\GeV^2} \Big( 
    \frac{C_T^{l 2}}{3.28^2} - \frac{C_{T5}^{l 2}}{1.89^2}
    - \frac{C_P^{l 2}}{10.36^2} - \frac{C_S^{l 2}}{5.98^2} \Big) + 
    \order{m_l^3} \Bigg] \cdot 10^{-7},
  \nonumber
\end{align} 
and the numerator of the normalized forward-backward asymmetry
\refeq{eq:Gaml:AFBl}
\begin{align} \nonumber
  \tau_{B^\pm}\, B_l = & \left[ \frac{\tau_{B^\pm}}{1.64 \rm ps} \right] 
  \Bigg[ 0.06 (C_S^l C_T^l + C_P^l C_{T5}^l) 
   + \frac{m_l}{\GeV} \Big(\frac{C_S^l}{6.25} - \frac{C_{T5}^l}{1.85} \Big)
\\
  & - \frac{m_l^2}{\GeV^2} \Big(\frac{C_S^l C_{T}^l}{4.12^2} + \frac{C_P^l
    C_{T5}^l}{4.12^2}\Big)  + \order{m_l^3} \Bigg] \cdot 10^{-7}.
  \label{eq:AFBnum:NP}
\end{align}
Here, we integrated over the dilepton mass region $1 \GeV^2 < q^2 \leq 7 \GeV^2$
and used the central values of the input parameters given in
\reftab{tab:num:input}.  Then $\FHl$ is given by the ratio of
\refeq{eq:FHlnum:NP} and \refeq{eq:BRl:NP}, $R_K$ by the ratio of
\refeq{eq:BRl:NP} for $l=\mu$ and $l=e$ and $\AFBl$ by the ratio of
\refeq{eq:AFBnum:NP} and \refeq{eq:BRl:NP}, respectively. The contributions of
the chirality flipped operators ${\cal{O}}^{l \prime}_{S,P}$ can be included by
the replacement $C_{S,P}^l \to C_{S,P}^l + C_{S,P}^{l \prime}$.

As can be seen from \refeq{eq:BRl:NP}, the $\BtoKll$ branching ratio is not very
sensitive to NP effects from scalar and tensor operators due to the small
coefficients in front of the NP couplings with respect to the large SM
contribution. Moreover, the SM uncertainties of $\BRl$ will hide NP unless the
Wilson coefficients become large, $C_i^{l \, \rm NP} \gtrsim 1$.  This actually
can happen in some NP scenarios as we will show, in particular, in the decays
into electrons, where the current experimental constraints are looser than the
ones for the muons. Due to its tiny theory uncertainty the ratio $R_K$ is a much
more powerful probe of NP than the $\BtoKll$ branching ratios.  Especially the
terms at zeroth order in the lepton mass but also the ones linear in $m_\mu$ can
significantly modify $R_K-1$ with respect to its negligible SM value.

The angular observables $\FHl$ \refeq{eq:FHlnum:NP} and $\AFBl$
\refeq{eq:AFBnum:NP} share several features with $R_K-1$: They have a small and
clean SM prediction and the sensitivity to tensor operators is higher than to
scalar and pseudoscalar ones. Note that the dependence of $\BRl$ and $\FHl$ on
the (pseudo-) scalar Wilson coefficients is the same and that the leading term
in the lepton-mass expansion of $\AFBl$ requires the presence of both (pseudo-)
scalar and tensor operators.  Note also that $R_K$ can be affected independently
by NP in $\BtoKee$ and $\BtoKmumu$ decays.

The available experimental information on $\FHl$, $R_K-1$ and $\AFBl$ including
SM predictions is given in \reftab{tab:comparison} together with other related
$b \to s \bar l l$ decay observables.  The data on $R_K$ include large dilepton
masses where QCDF is not applicable and the ones on $\FHl$ and $\AFBl$ are in
addition lepton flavor averaged. We do not take these constraints into account
since they cannot be applied in a straightforward way besides having sizeable
uncertainties.

%\begin{table}
\TABLE[ht]{
\renewcommand{\arraystretch}{1.1}
\begin{tabular}{||l|c|c|c||}
\hline \hline
  observable & sensitive to & SM value& data\\
  \hline \hline
  $\FHmu$ & $C_{S,P}^\mu +C_{S,P}^{\mu  \prime}$, $C_{T(5)}^\mu$  &
  ${\cal{O}}(m_\mu^2/q^2)$ & 
  $0.81^{+0.58}_{-0.61} \pm 0.46^\dagger$ \cite{Aubert:2006vb} 
\\
\hline
  $A_{\rm FB}^\mu$ & $C_{S,P}^\mu +C_{S,P}^{\mu \prime}$, $C_{T(5)}^\mu$ & 
  ${\cal{O}}(\alpha_e/(4 \pi))$ & 
  $0.15^{+0.21}_{-0.23} \pm 0.08^\dagger$ \cite{Aubert:2006vb} \\
\mbox{} & \mbox{} & \mbox{}&  $0.10 \pm 0.14 \pm 0.01^\dagger$  \cite{Ishikawa:2006fh}
\\ \hline
  $R_K-1$ & $C_{S,P}^l + C_{S,P}^{l \prime}$, $C_{T(5)}^l$, $e$ vs.~$\mu$ &
  ${\cal{O}}(10^{-4})$ & 
  $0.24 \pm 0.31^\dagger$ \cite{Abe:2004ir, Aubert:2006vb}
\\ \hline \hline
  ${\cal{B}}(\bar B_s \to \bar \mu \mu )$ & $C_{S,P}^\mu -C_{S,P}^{\mu \prime}$ & 
  $(3.23 \pm 0.44) \cdot 10^{-9}$ & 
  $<8.0 \cdot 10^{-8}$ \cite{Scuri:2007py} \\
\hline
  ${\cal{B}}(\bar B_s \to \bar e e )$ & $C_{S,P}^e -C_{S,P}^{e \prime}$ & 
  $(7.56 \pm 0.32) \cdot 10^{-14}$ & 
  $<5.4 \cdot 10^{-5}$ \cite{Acciarri:1996us} \\
\hline
$\BRincl{\mu}{>0.04}{ \mbox{}}$ 
& $C_{S}^{\mu (\prime)} \pm C_{P}^{\mu  (\prime)}$, $C_{T(5)}^\mu$ & 
$(4.15 \pm 0.70) \cdot 10^{-6}$ \cite{Ali:1999mm}
& $ (4.3 \pm 1.2) \cdot 10^{-6}$ \cite{Yao:2006px}\\
\hline
$\BRincl{e}{>0.04}{ \mbox{}}$ 
& $C_{S}^{e (\prime)} \pm C_{P}^{e  (\prime)}$, $C_{T(5)}^e$ & 
$(4.15 \pm 0.70) \cdot 10^{-6}$ \cite{Ali:1999mm} 
& $ (4.7 \pm 1.3) \cdot 10^{-6}$ \cite{Yao:2006px}\\
\hline \hline
\end{tabular}
\caption{\label{tab:comparison} Observables in $b \to s \bar l l$ induced 
transitions. Upper bounds are given at $90 \%$ C.L. For details see text.
$^\dagger$Data include $q^2$-regions where QCDF does not apply and both $l=e$ 
and $\mu$ are included.}}
%\end{table}

Important constraints on NP come from $\BR(\Bstoll)$, which can be written as
\begin{align}
  \label{eq:BR:Bsll}
  \BR(\Bstoll) & = \frac{\GF^2 \alE^2 M^5_{B_s} f_{B_s}^2 \tau_{B_s}}{64 \pi^3}
     |V_{tb}^{}V_{ts}^{\ast}|^2 \sqrt{1 - \frac{4 m_l^2}{M_{B_s}^2}} 
\nonumber\\
  & \times \Bigg\{ 
    \Bigg(1 - \frac{4m_l^2}{M_{B_s}^2} \Bigg) \Bigg|
\frac{C_S^l - C^{l \prime}_S}{m_b + m_s}\Bigg|^2 
    + \Bigg|\frac{C_P^l - C^{l  \prime}_P}{m_b + m_s} + \frac{2 m_l}{M^2_{B_s}} C_{10}\Bigg|^2 \Bigg\}.
\end{align}
The $\Bstoll$ branching ratios depend on the difference of Wilson coefficients
$(C_{S,P}^l - C_{S,P}^{l \prime})$. It follows that constraints from
\refeq{eq:BR:Bsll} can be evaded in the presence of both unprimed and primed
(pseudo-)~scalar Wilson coefficients unless there is a complementary constraint
such as on $(C_{S,P}^l + C_{S,P}^{l \prime})$ from $\BtoKll$ decays
\cite{Hiller:2003js}. Tensor operators do not contribute to $\Bstoll$ decays and
hence $C_{T,T5}^l$ are not constrained by these decays.  The current $90\%$
C.L.~upper bound on $\BR(\Bstoll)$ for $l=\mu$ comes from CDF and \D0
\cite{Scuri:2007py}\footnote{ A stronger bound has been reported from a combined
  CDF and \D0 analysis at $95 \%$ C.L., $\BR(\Bstomm) < 5.8 \cdot 10^{-8}$
  \cite{Maciel:Bsmm}.}  and for electrons from L3 \cite{Acciarri:1996us}.  The
experimental information can be seen in \reftab{tab:comparison} together with
the SM predictions obtained with the input from \reftab{tab:num:input}.  The
bound on $\BR(\Bstomm)$ is ${\cal{O}}(20)$ away from the SM, and the one for
electrons is nine orders of magnitude above the SM. As we show in
\refsec{sec:scen1}, the current $\BR(\Bstoee)$ constraint is nevertheless on the
verge of being useful, since NP in $C_{S,P}^{l (\prime)}$ does not enter the
$\Bstoll$ modes with $m_l$-suppression as the SM contribution, see
\refeq{eq:BR:Bsll}.

We further take into account the measurements of the branching ratios of the
inclusive $\bar B \to X_s \bar e e$ and $\bar B \to X_s \bar \mu \mu$ decays for
$q^2 >0.04 \GeV^2$ denoted by $\BRincl{l}{>0.04}{ \mbox{}}$.  The corresponding
experimental values \cite{Yao:2006px,Aubert:2004it} can be seen in
\reftab{tab:comparison} with SM predictions from \cite{Ali:1999mm}.  The
$q^2$-cut dependent $\BXsll$ branching ratios with (pseudo-) scalar and tensor
interactions can be written as (see, e.g., \cite{Fukae:1998qy})
\begin{align} \label{eq:BR:BXsll}
   \BRincl{l}{q^2_{\rm min},\, q^2_{\rm max}}{} & \equiv \BR(\BXsll)  =
  \,\,\BRincl{l}{q^2_{\rm min},\, q^2_{\rm max}}{,{\rm SM}} + ( |C^l_T|^2 + |C^l_{T5}|^2) 
{\cal M}_T  \\[2mm]  
   & + (|C^l_S + C^l_P|^2 + |C^{l  \prime}_S + C^{l \prime}_P|^2 
 + |C^l_S - C^l_P|^2 + |C^{l  \prime}_S - C^{l \prime}_P|^2) {\cal M}_S, 
\nonumber
\end{align}
where
\begin{align} \label{eq:calMST} 
  {\cal M}_{S,T} & = \frac{\BR_0}{2 m_b^8} \int_{q^2_{\rm min}}^{q^2_{\rm max}} dq^2\; M_{S,T}(q^2), &
  \BR_0 & = \frac{3 \alE^2}{(4 \pi)^2} \frac{|V_{tb}^{}V_{ts}^{\ast}|^2}{|V_{cb}|^2} 
            \frac{\BR(\BXclv)}{f(m_c/m_b) \kappa(m_c/m_b)} 
\end{align}
and
\begin{align} \label{eq:MST}
   M_S(q^2) & = 2 q^2 (m_b^2-q^2)^2,  
 & M_T(q^2) & = \frac{64}{3} (m_b^2-q^2)^2 (2 m_b^2+q^2).
\end{align}
Here we neglect kinematical factors of $m_s$ and $m_l$ in the NP part and
evaluate \refeq{eq:calMST} and \refeq{eq:MST} with a $b$-quark mass of $4.8
\GeV$, corresponding to the pole mass in accordance with \cite{Ali:1999mm}. The
functions $f(m_c/m_b)$ and $\kappa(m_c/m_b)$ represent the phase space function
and QCD corrections of the decay $\BXclv$, respectively, and can be seen in
\cite{Fukae:1998qy}. Since $M_{S,T} >0$, NP from (pseudo-) scalar and tensor
contributions enhances the $\BXsll$ branching ratios, and only the upper
boundary of the experimental value of $\BR(\BXsll)$ becomes a constraint on the
corresponding Wilson coefficients.  Also, since $M_T \gg M_S$, the inclusive
branching ratios are more sensitive to tensor than scalar and pseudoscalar
operators.  Numerically, for $0.04 \GeV^2 < q^2 \leq m_b^2$ we obtain ${\cal
  M}_{S}=1.92 \cdot 10^{-8}$ and ${\cal M}_{T}=1.84 \cdot 10^{-6}$.

In our NP analysis we also predict $\BRincl{l}{1,6}{}$ for $l=e,\mu$ using
$\BRincl{e}{1,6}{,{\rm SM}} = (1.64 \pm 0.11) \cdot 10^{-6}$ and
$\BRincl{\mu}{1,6}{,{\rm SM}} = (1.59 \pm 0.11) \cdot 10^{-6}$
\cite{Huber:2005ig}. These values are close to the experimental world average
$\BRincl{l}{1,6}{,{\rm exp}} = (1.60 \pm 0.51) \cdot 10^{-6}$
\cite{Aubert:2004it} which is lepton flavor averaged and we therefore do not
consider it as a constraint. However, we use this to illustrate the physics
potential of future lepton flavor specific $\BRincl{e}{1,6}{}$ and
$\BRincl{\mu}{1,6}{}$ measurements.  The ${\cal M}_{S,T}$-coefficients for this
low dilepton mass region $1 \GeV^2 < q^2 \leq 6 \GeV^2$ are ${\cal M}_{S}=0.52
\cdot 10^{-8}$ and ${\cal M}_{T}=0.83 \cdot 10^{-6}$. Note that we used here the
$b$-quark pole mass in the NP part of $\BR(\BXsll)$ as well. To be consistent
with the SM results of \cite{Huber:2005ig} the $1S$ mass should be used once the
next-to-leading order corrections to the NP part are known.

Given the existing experimental constraints we cannot perform at present a fully
model-independent analysis and fit for the six real NP Wilson coefficients per
lepton species. Instead, we entertain in the following four benchmark scenarios
with (pseudo-) scalar operators (Scenario I-III) and the tensor operators
(Scenario IV) defined as:
\begin{itemize}
\item[--] Scenario I: NP in $C_S^l$ and $C_P^l$, all other NP contributions
  vanish.
\item[--] Scenario II: Same as Scenario I, but with the additional assumptions
  $C_S^l = - C_P^l$ and $C^l \propto m_l$.
\item[--] Scenario III: NP in $C_S^l$, $C_P^l$ and $C_S^{l \prime}$, $C_P^{l
    \prime}$, the tensor coefficients $C_{T,T5}^l$ vanish.
\item[--] Scenario IV: NP in the tensor coefficients $C_T^l$, $C_{T5}^l$, all
  other NP contributions vanish.
\end{itemize}
Scenario II is inspired by the Minimal Supersymmetric Standard Model (MSSM) for
large values of $\tan \beta$, where $\tan \beta$ denotes the ratio of the two
Higgs vacuum expectation values, see also \refsec{sec:tensors}.

All $\BtoKll$ observables are obtained for $1 \GeV^2 < q^2 \leq 7 \GeV^2$.  We
employ all bounds at $90 \%$ C.L.  The resulting allowed ranges of the NP Wilson
coefficients and the $b \to s \bar l l$ decay observables are summarized in
\reftab{tab:NP:wc} and \reftab{tab:NP:res}, respectively. Since the current
experimental errors dominate the uncertainties, in the following we do not take
into account SM uncertainties. Their inclusion would allow for slightly bigger
NP effects.

%
%----------------------------------------------------
\subsection{Scenario I: Scalars $C_S^l$ and $C_P^l$ \label{sec:scen1}}

\begin{figure}
\begin{center}
  \epsfig{figure=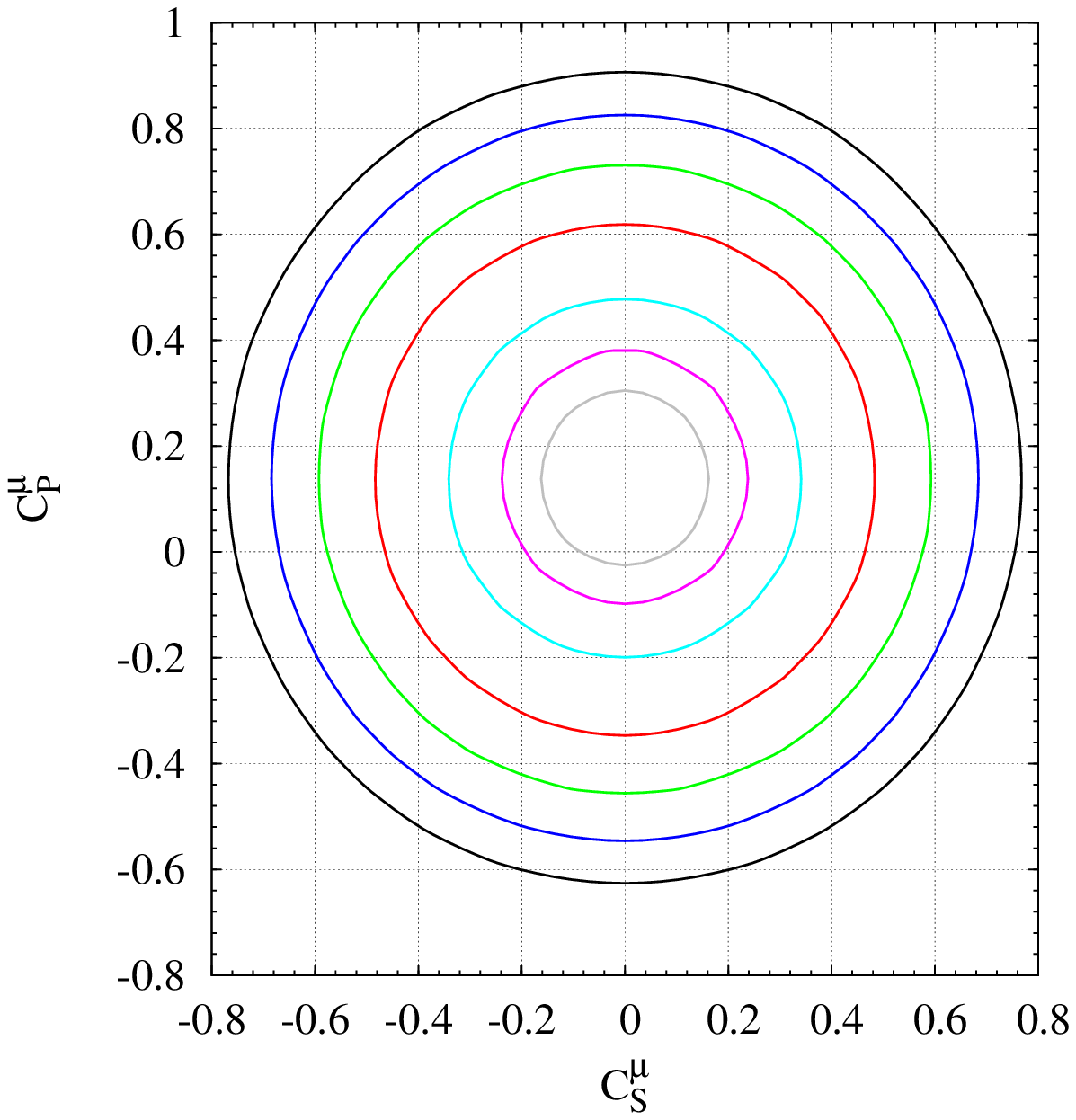,     height=7.5cm, angle=0} 
  \epsfig{figure=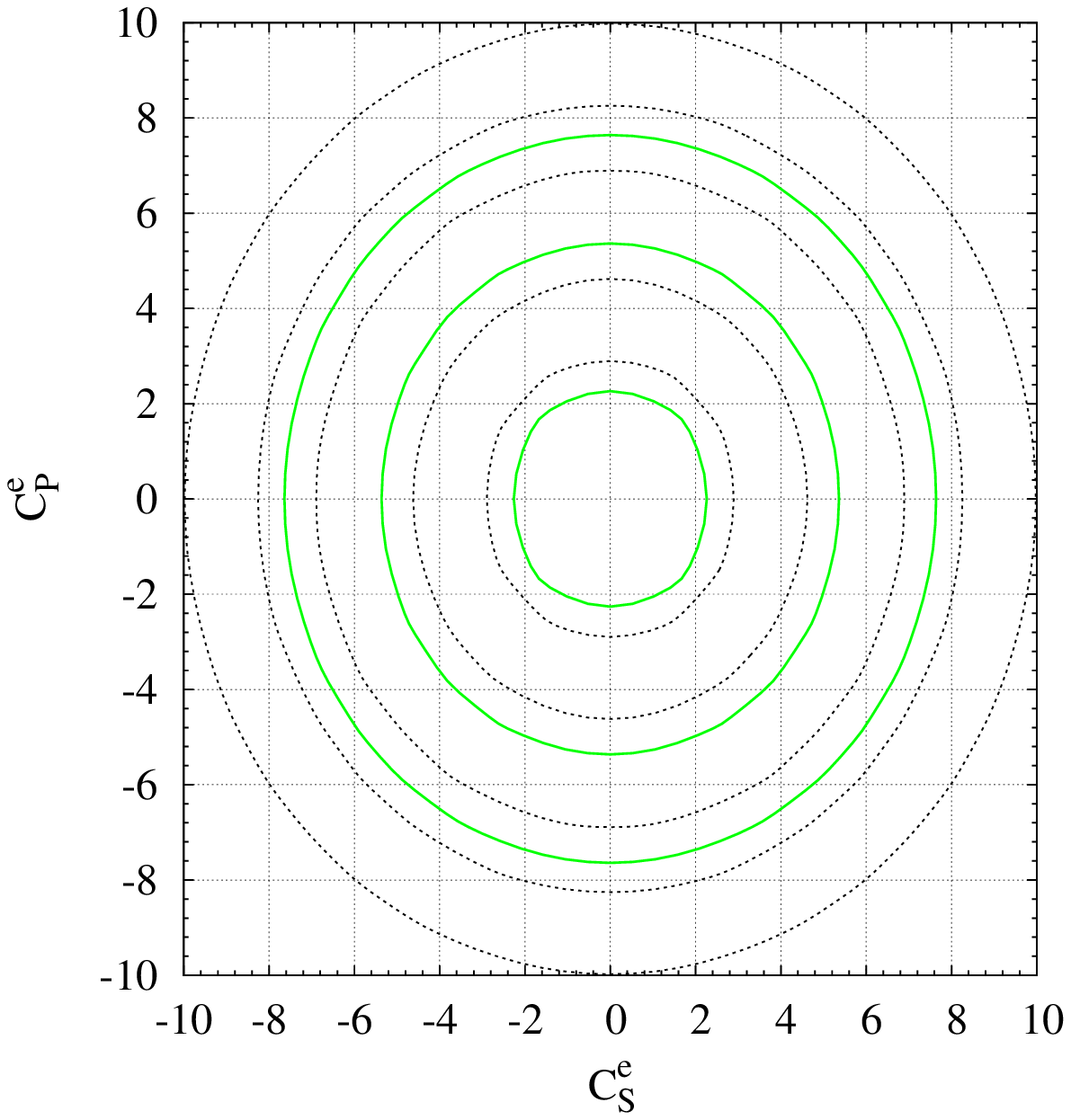,     height=7.5cm, angle=0} 
\end{center}
\caption{ \label{fig:scen:I} In the left-hand plot contours of $\BR(\Bstomm)$
  are shown in the $C^\mu_S - C^\mu_P$ plane in Scenario I. The contours enclose
  values of $\BR(\Bstomm) < \{0.05, 0.1, 0.2, 0.4, 0.6, 0.8, 1.0\} \cdot
  10^{-7}$ starting with the innermost. In the right-hand plot contours of
  $\BRincl{e}{>0.04}{} < \{4.5, 5.0, 6.0, 6.8, 8.0\} \cdot 10^{-6}$ (dashed
  black) and $\BR(\Bstoee) < \{0.1, 0.5, 1.0\} \cdot 10^{-5} $ (solid green) are
  shown in the $C^e_S - C^e_P$ plane in Scenario I starting with the innermost.}
\end{figure}

We start with a discussion of the Wilson coefficients for muons, $C_{S,P}^\mu$.
The bounds on $C_{S,P}^\mu$ from \refeq{eq:BR:Bsll} are displayed in the left
hand plot of \reffig{fig:scen:I}, where contours of $\BR(\Bstomm) < \{0.05, 0.1,
0.2, 0.4, 0.6, 0.8, 1.0\} \cdot 10^{-7}$ are shown. The ranges for $C_{S,P}^\mu$
after applying the $90 \%$ C.L.  $\BR(\Bstomm)$ upper bound given in
\reftab{tab:comparison} can be seen in \reftab{tab:NP:wc}. The corresponding
ranges of the observables are presented in \reftab{tab:NP:res}. As can be seen,
$\FHmu$ can deviate from the SM by about $40\%$ whereas the forward-backward
asymmetry is less then $1\%$ in agreement with and updating earlier findings
\cite{Bobeth:2001sq}. The deviation of the branching ratio $\BRmu$ from the SM
is less than $2\%$ and completely negligible in view of the theoretical
uncertainties. Also the NP contributions to $\BRincl{\mu}{1,6}{}$ and
$\BRincl{\mu}{>0.04}{}$ are small compared to the theoretical uncertainties.

\begin{figure}
\begin{center}
  \epsfig{figure=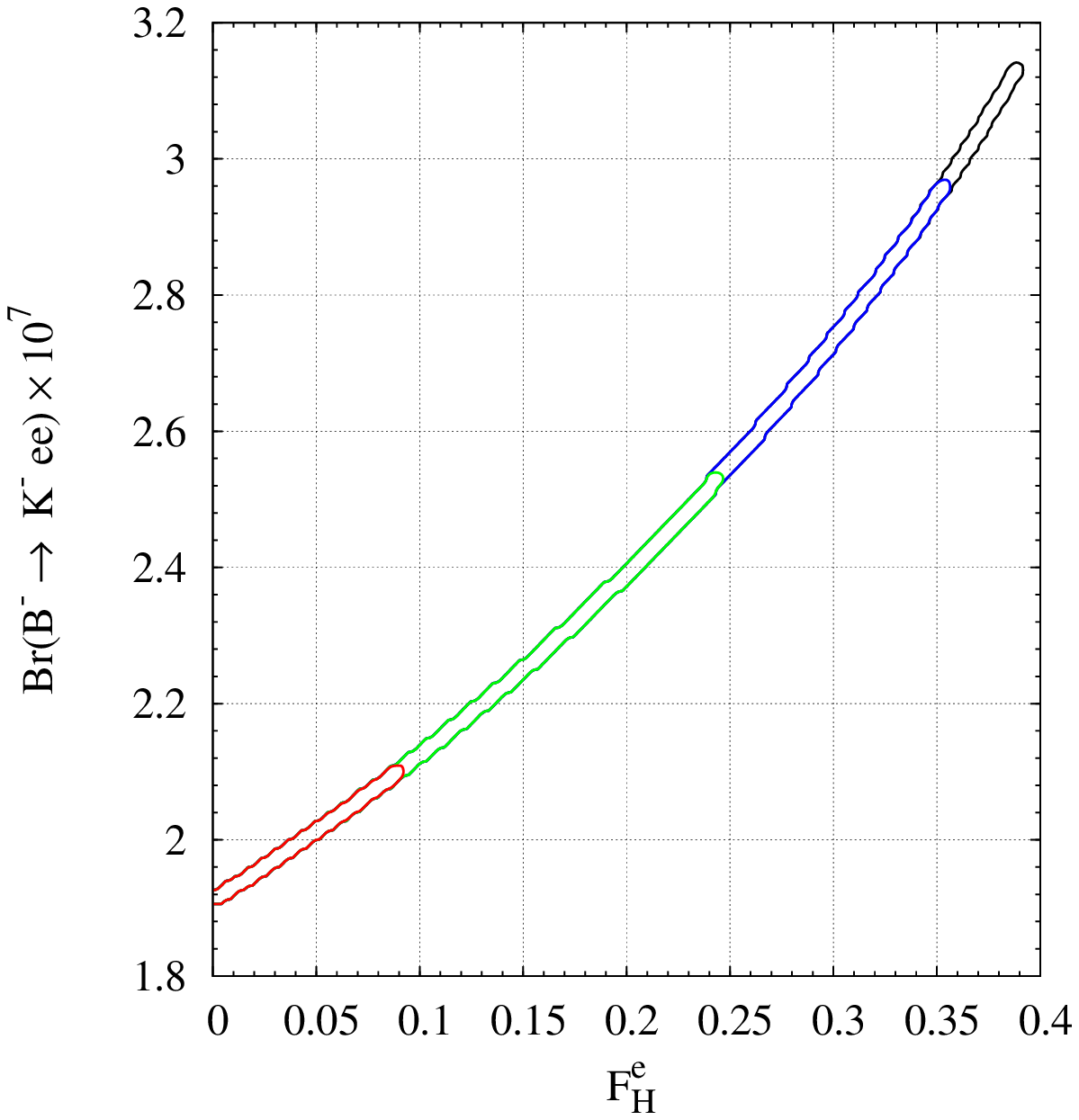,   height=7.5cm, angle=0} 
  \epsfig{figure=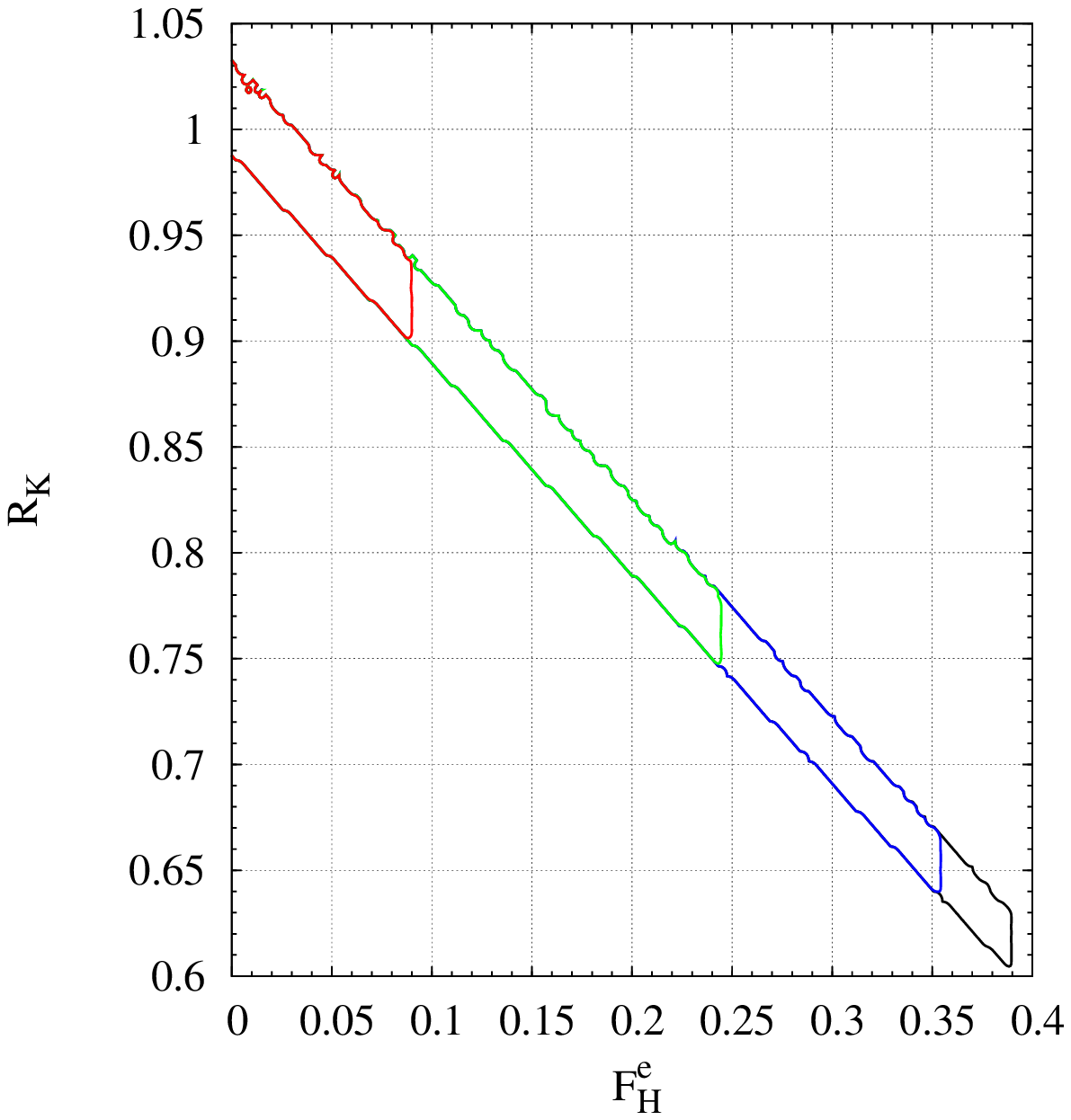,     height=7.5cm, angle=0} 
\end{center}
\caption{ \label{fig:sc1:FHe} Contours of $\BRincl{e}{1,6}{} < \{1.75, 2.0,
  2.25, 2.35 \} \cdot 10^{-6}$ in the $\FHe - \BRe$ plane (left-hand plot) and
  the $\FHe - R_K$ plane (right-hand plot) in Scenario I.}
\end{figure}

The situation for the electrons is different due to the weaker bound from
$\BR(\Bstoee)$ such that $\BRincl{e}{>0.04}{}$ gives the strongest constraint on
$C_{S,P}^e$. In the right-hand plot of \reffig{fig:scen:I} contours in the
$C^e_S -C^e_P$ plane are shown for $\BRincl{e}{>0.04}{} < \{4.5, 5.0, 6.0, 6.8,
8.0\} \cdot 10^{-6}$ and $\BR(\Bstoee) < \{0.1, 0.5, 1.0\} \cdot 10^{-5} $. The
latter illustrates the constraints of improved measurements of $\BR(\Bstoee)$.
We encounter the large ranges of $C^e_{S,P}$ given in \reftab{tab:NP:wc} allowed
by $\BRincl{e}{>0.04}{} <6.8 \cdot 10^{-6}$ at $90\%$ C.L., see
\reftab{tab:comparison}. The corresponding ranges for the decay observables for
$l = e$ are presented in \reftab{tab:NP:res}. As one can see, $\FHe$ can be
enhanced by orders of magnitude compared to its negligible SM value.
Furthermore, $\FHe$ is strongly correlated to $\BRe$, $R_K$ (see
\refeq{eq:BRl:NP} and \refeq{eq:FHlnum:NP}) and $\BRincl{e}{1,6}{}$.  The
observables $\FHe$, $\BRe$ and $\BRincl{e}{1,6}{}$ increase for increasing
$|C_{S,P}^e|$ whereas $R_K$ decreases.  We show $\BRe$ versus $\FHe$ (left-hand
plot) and $R_K$ versus $\FHe$ (right-hand plot) in \reffig{fig:sc1:FHe}.  The
branching ratio $\BRe$ can be enhanced by about $60\%$ with respect to its SM
value.  $\BRincl{e}{1,6}{}$ exhibits a similar enhancement but is subject to
smaller theoretical uncertainties. The forward-backward asymmetry $\AFBe$ is
negligibly small.

The observable $R_K$ depends on both lepton channels $l = e, \mu$. In Scenario I
the denominator $\BRe$ receives large NP contributions whereas the numerator
$\BRmu$ stays close to its SM value due to the strong constraint from $\Bstomm$.
This leads to a substantial decrease of $R_K$ with respect to the SM as can be
seen in \reftab{tab:NP:res} and also in the right-hand plot of
\reffig{fig:sc1:FHe}.

\TABLE[ht]{
\begin{tabular}{|| c | c  c  c  c ||}
\hline \hline
 Wilson coefficient   & Sc I & Sc II & Sc III & Sc IV 
\\
\hline \hline
$C_{S,P}^e$    & $[-8.3, 8.3]$   & $-$             & $[-8.3,8.3]$ & $-$
\\[0.5ex]
$C_S^\mu$      & $[-0.69, 0.69]$ & $[-0.55, 0.41]$ & $[-5.6,5.6]$ & $-$
\\[0.5ex]
$C_P^\mu$      & $[-0.55, 0.82]$ & $ = -C_S^\mu$    & $[-5.6,5.6]$ & $-$
\\[0.5ex] 
$C_{S,P}^{e \prime}$  & $-$             & $-$             & $[-8.3,8.3]$ & $-$
\\[0.5ex]
$C_{S,P}^{\mu \prime}$ & $-$            & $-$             & $[-5.6,5.6]$ & $-$
\\[0.5ex]
$C_{T,T5}^e$    & $-$             & $-$             & $-$          & $[-1.2, 1.2]$
\\[0.5ex]
$C_{T,T5}^\mu$  & $-$             & $-$             & $-$          & $[-1.1, 1.1]$ 
\\[0.5ex]
\hline \hline 
\end{tabular}
\caption{ \label{tab:NP:wc} The allowed ranges for the NP Wilson coefficients
  $C_i^l$ in Scenarios I-IV after using the constraints $\BR(\Bstoee) < 5.4
  \cdot 10^{-5}$, $\BR(\Bstomm) < 0.8 \cdot 10^{-7}$, $\BR^{\rm incl}_e|_{>0.04}
  < 6.8 \cdot 10^{-6}$ and $\BR^{\rm incl}_\mu|_{>0.04} < 6.3 \cdot 10^{-6}$,
  see \reftab{tab:comparison}. A ``$-$'' means that the corresponding
  coefficient is zero in this NP scenario. }}

\TABLE[ht]{
\begin{tabular}{|| c || c | c | c | c ||}
\hline \hline
Observable & Sc I & Sc II & Sc III & Sc IV 
\\
\hline \hline
$\FHe$  & $< 0.39$          & $-$                & $< 0.56$         & $ < 0.13$
\\[0.5ex]
$\FHmu$ & $[0.013, 0.035]$ & $[0.018, 0.032]$ & $[0.013, 0.56]$ & $[0.014, 0.18]$
\\[0.5ex]
$R_K$   & $[0.61, 1.01]$    & $[0.996, 1.01]$ & $[0.44, 2.21]$ & $[0.93, 1.10]$
\\[0.5ex]
%$R_K |_{e = {\rm SM}}$ & $[0.991, 1.013]$ & $-$              & $[0.991, 2.23]$ %& $[0.977, 1.099]$
%\\[0.5ex]
$\BRe \, [10^{-7}]$  & $[1.91, 3.14]$ & $-$            & $[1.91, 4.36]$ & $[1.91, 2.00]$
\\[0.5ex]
$\BRmu \, [10^{-7}]$ & $[1.90, 1.94]$ & $[1.90, 1.93]$ & $[1.90, 4.26]$ & $[1.87, 2.10]$
\\[0.5ex]
$\AFBe \, [\%] $ & $[-0.02, 0.02]$ & $-$           & $[-0.02, 0.02]$ & $[-0.02, 0.02]$
\\[0.5ex]
$\AFBmu \, [\%] $ & $[-0.6, 0.6]$   & $[-0.5, 0.3]$ & $[-4.46, 4.46]$ & $[-3.1, 3.1]$
\\[0.5ex]
\hline
$\BR(\Bstoee) \, [10^{-5}]$ & $< 1.17$ & $-$     & $< 2.33$ & $-$
\\[0.5ex]
$\BR(\Bstomm) \, [10^{-7}]$ & $< 0.8$  & $< 0.8$ & $< 0.8$  & $-$
\\[0.5ex]
$\BRincl{e}{1,6}{} \, [10^{-6}]$ & $[1.64, 2.35]$ & $-$            & $[1.64, 2.35]$ & $[1.64, 2.83]$
\\[0.5ex]
$\BRincl{\mu}{1,6}{} \, [10^{-6}]$ & $[1.59, 1.60]$ & $[1.59, 1.60]$ & $[1.59, 2.17]$ & $[1.59, 2.56]$
\\[0.5ex]
$\BRincl{e}{>0.04}{}   \, [10^{-6}]$ & $[4.15 , 6.8]$ & $-$            & $[4.15, 6.8]$  & $[4.15, 6.8]$
\\[0.5ex]
$\BRincl{\mu}{>0.04}{} \, [10^{-6}]$ & $[4.15, 4.18]$ & $[4.15, 4.17]$ & $[4.15, 6.3]$  & $[4.15, 6.3]$
\\[0.5ex]
\hline \hline 
\end{tabular}
\caption{ \label{tab:NP:res} Allowed ranges for $b \to s \bar l l$ observables
  in Scenarios I-IV after taking into account the constraints from
  $\BR(\Bstoll)$ and $\BRincl{l}{>0.04}{}$ for $l=e$ and $l=\mu$, see
  \reftab{tab:comparison} and the text for details. A ``$-$'' means that the
  corresponding observable is SM-like.  }}

%
%----------------------------------------------------
\subsection{Scenario II: MSSM-like $C_S^\mu =-C_P^\mu$}

Scenario II is a special case of Scenario I inspired by the MSSM in a certain
limit (large $\tan \beta$), see also \refsec{sec:tensors}. In this model, the
(pseudo-) scalar Wilson coefficients are proportional to the lepton mass
$C_{S,P}^l \sim m_l$, such that $C_{S,P}^e$ can be neglected and $b \to s \bar e
e$ decays are SM-like. Furthermore the relation $C^\mu_S =-C^\mu_P$ holds and
the primed coefficients $C_{S,P}^{\mu \prime}$ are suppressed by $m_s/m_b$ and
can be neglected.

The allowed range of $C^\mu_S$ and the effects of NP on the rare decay
observables are given in \reftab{tab:NP:wc} and \reftab{tab:NP:res},
respectively. Since Scenario II is a constrained variant of Scenario I the
deviations from the SM are smaller in the former.  The NP contributions to
$\FHmu$ do not exceed $30\%$ whereas the deviations of $\BRmu$ from the SM are
of the order of $2\%$, much smaller than the theoretical uncertainties.  The
same holds for $\BRincl{\mu}{1,6}{}$, which confirms earlier studies within the
MSSM \cite{Chankowski:2003wz}. Since $\BRe$ is SM-like in Scenario II, the
deviation of $R_K$ from the SM is much reduced with respect to the one in
Scenario I. We find NP effects of $1\%$, which are larger than the uncertainties
of the SM prediction. The forward-backward asymmetry is smaller then $1\%$ in
agreement with previous works in the framework of the MSSM \cite{Demir:2002cj}.

%
%----------------------------------------------------
\subsection{Scenario III: Scalars $C_S^l$, $C_P^l$ and $C_S^{l \prime}$, 
$C_P^{l \prime}$ }

In Scenario III we use the full set of (pseudo-) scalar Wilson coefficients
including the chirality flipped ones $C^{l \prime}_{S,P}$ for $l =e$ and
$l=\mu$. The constraint from the $\Bstoll$ branching ratios alone can be evaded
due to cancellations between $C^{l}_{S,P}$ and $C^{l \prime}_{S,P}$, see
\refeq{eq:BR:Bsll}. To obtain constraints on $C^{l (\prime)}_{S,P}$ we combine
$\BR(\Bstoll)$ with $\BRincl{l}{>0.04}{}$ data. We find the allowed ranges for
the Wilson coefficients given in \reftab{tab:NP:wc}. In the electron sector
$C_{S,P}^e$ can be as big as in Scenario I with identical ranges for $C_{S,P}^{e
  \prime}$. In the muon sector the Wilson coefficients $C^{\mu (\prime)}_{S,P}$
are now comparable in magnitude to the ones for electrons.

The large Wilson coefficients lead to big NP effects in the rare decay
observables, see \reftab{tab:NP:res}. In Scenario III $R_K$ can both increase
and decrease significantly with respect to the SM as opposed to Scenario I where
$\BR(\Bstomm)$ permits only a large decrease of $R_K$.  The substantial
deviations of $R_K$ from the SM are already challenged by existing data given in
\reftab{tab:comparison}.  However, as already stressed, since these data contain
also large-$q^2$ events where QCDF is not applicable it is not clear how to
impose these constraints in a well-defined way.

\begin{figure}
\begin{center}
  \epsfig{figure=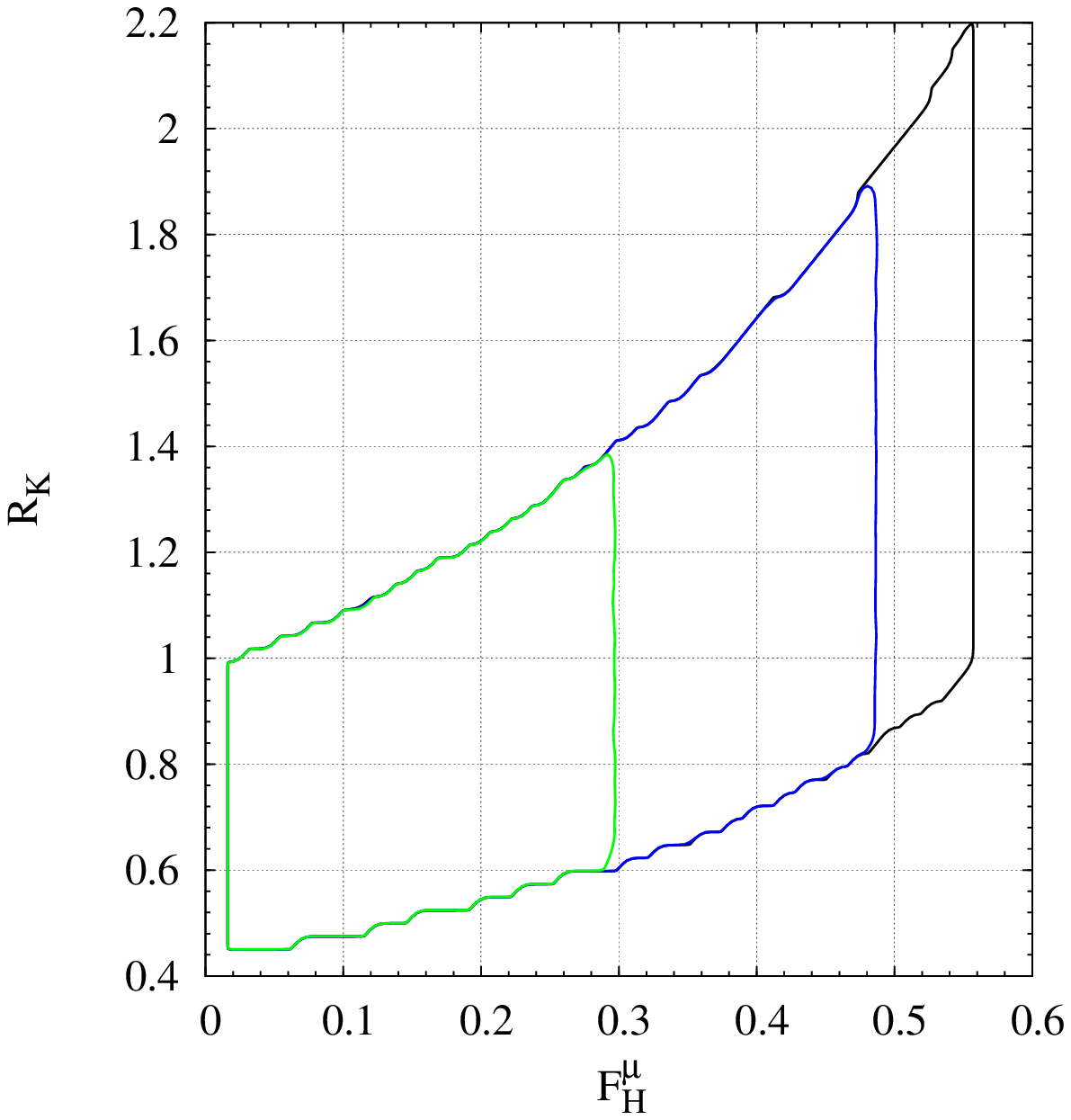,  height=7.5cm, angle=0} 
  \epsfig{figure=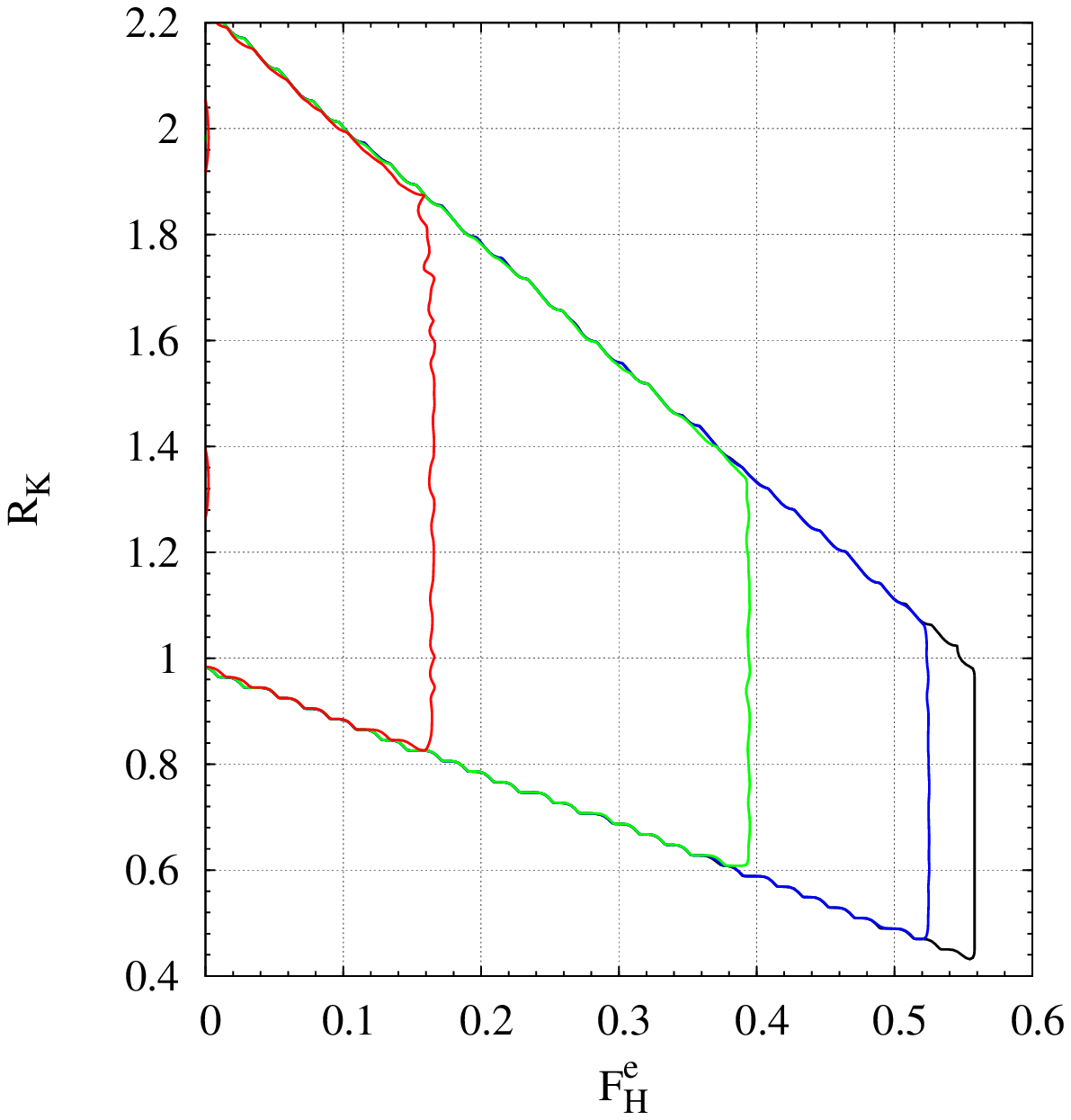,  height=7.5cm, angle=0} 
\end{center}
\caption{ \label{fig:sc3} Contours of $\BRincl{\mu}{1,6}{} < \{1.75, 2.0, 2.17
  \} \cdot 10^{-6}$ in the $\FHmu - R_K$ plane in Scenario III (left-hand plot).
  In the right-hand plot contours of $\BRincl{e}{1,6}{} < \{1.75, 2.0, 2.25,
  2.35 \} \cdot 10^{-6}$ are shown in the $\FHe - R_K$ plane in Scenario III.
  For details see text. }
\end{figure}

The increase of both $\FHmu$ and $R_K$ for increasing values of
$\BRincl{\mu}{1,6}{}$ can be seen in the left-hand plot of \reffig{fig:sc3},
where contours of $\BRincl{\mu}{1,6}{} < \{1.75, 2.0, 2.17 \} \cdot 10^{-6} $
are shown.  Similarly, the increase of $\FHe$ and decrease of $R_K$ for
increasing values of $\BRincl{e}{1,6}{}$ is displayed in the right-hand plot of
\reffig{fig:sc3} with contours of $\BRincl{e}{1,6}{} < \{1.75, 2.0, 2.25, 2.35
\} \cdot 10^{-6}$. The NP contributions enhance both $\BRe$ and $\BRmu$ by order
$200\%$ above the SM such that measurements of these observables in the
low-$q^2$ region could provide constraints regardless of the large form factor
uncertainties. Scenario III allows for $|\AFBmu| \lesssim (4-5)\%$ whereas
$\AFBe$ is negligibly small.

%
%----------------------------------------------------
\subsection{Scenario IV: Tensors $C_T^l$, $C_{T5}^l$ \label{sec:scen4}}

\begin{figure}
\begin{center}
  \epsfig{figure=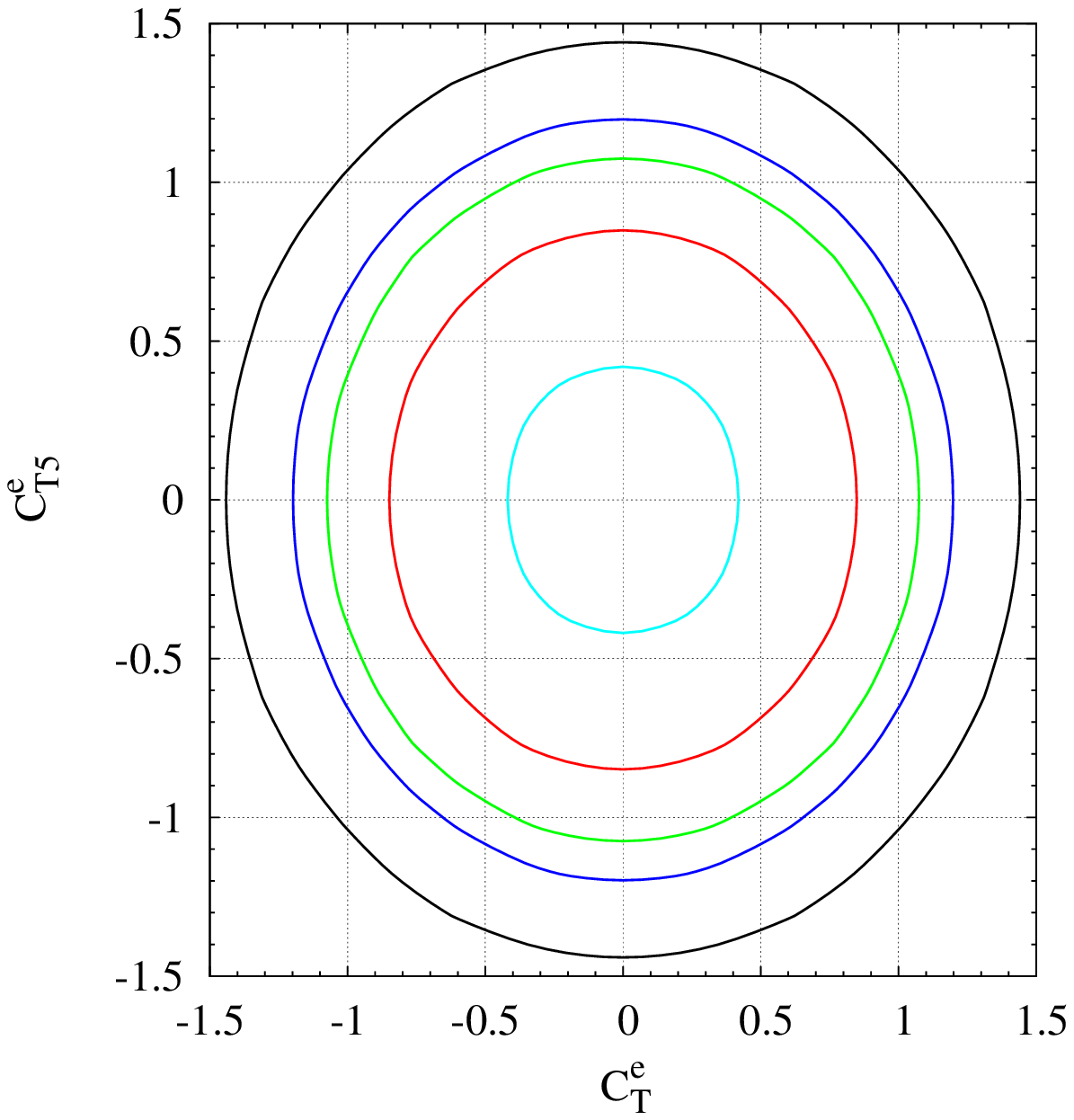,         height=7.5cm, angle=0} 
  \epsfig{figure=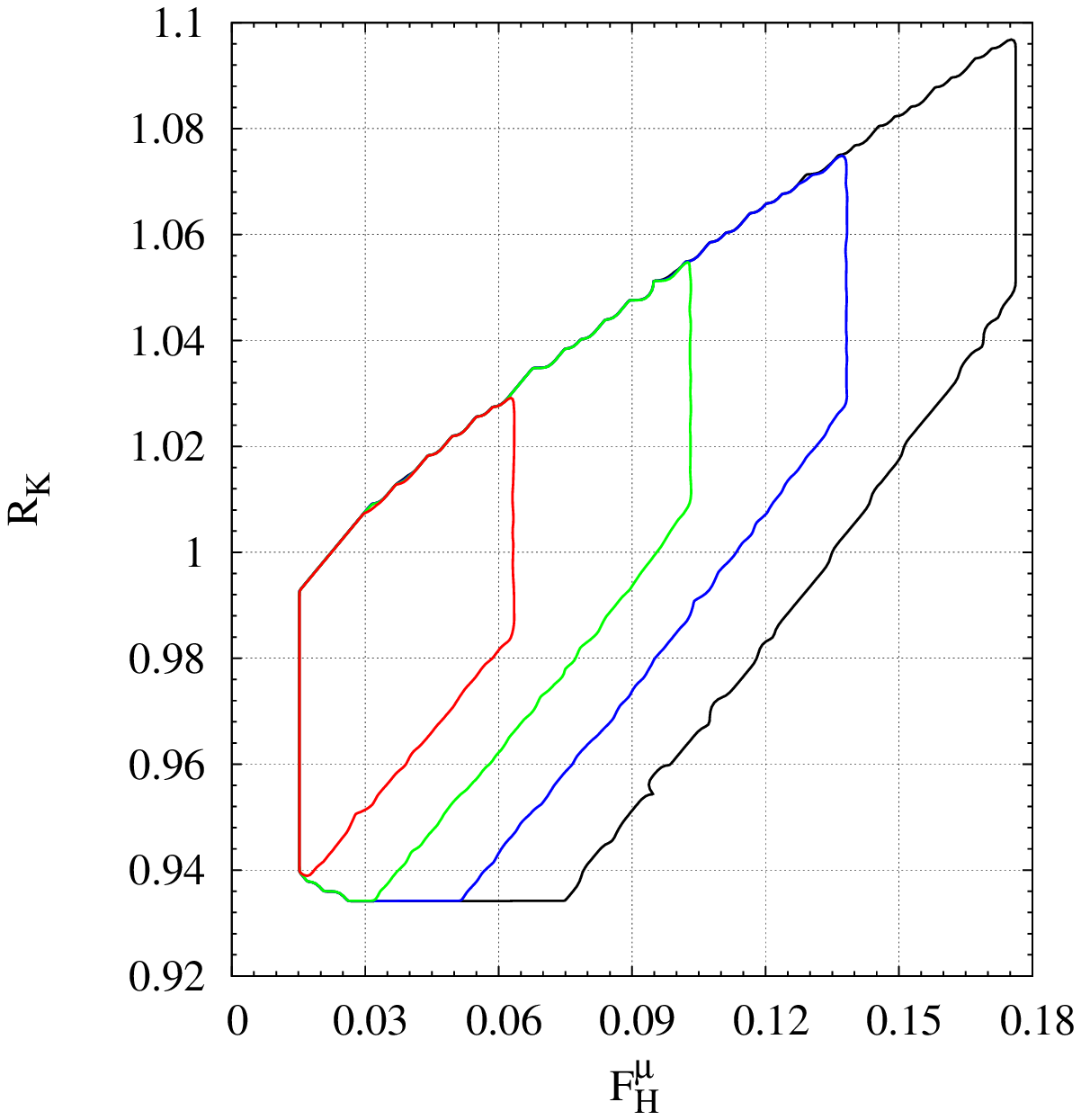,  height=7.5cm, angle=0} 
\end{center}
\caption{ \label{fig:sc4} In the left-hand plot contours of
  $\BRincl{e}{>0.04}{}$ are shown in the $C^e_T - C^e_{T5}$ plane in Scenario
  IV. Each contour encloses values of $\BRincl{e}{>0.04}{} < \{4.5, 5.5, 6.3,
  6.8, 8.0\} \cdot 10^{-6}$ starting with the innermost. Corresponding
  constraints for $C^\mu_T - C^\mu_{T5}$ can be read off from the left-hand plot
  as well.  In the right-hand plot contours of $\BRincl{\mu}{1,6}{} < \{1.75,
  2.0, 2.25, 2.56\} \cdot 10^{-6}$ are shown for $\FHmu$ versus $R_K$ in
  Scenario IV starting with the innermost.}
\end{figure}

In Scenario IV we consider only NP in $C^l_{T, T5}$. These Wilson coefficients
do not contribute to $\Bstoll$ decays and hence are currently constrained only
by inclusive $\BXsll$ decays \refeq{eq:BR:BXsll}. The corresponding bounds can
be seen in the left-hand plot of \reffig{fig:sc4}, where contours of
$\BRincl{e}{>0.04}{} < \{4.5, 5.5, 6.3, 6.8, 8.0\} \cdot 10^{-6}$ are shown in
the $C^e_T - C^e_{T5}$ plane starting with the innermost.  The constraints on
$C^\mu_{T, T5}$ from upper bounds on $\BRincl{\mu}{>0.04}{}$ can be read off
from the same plot. We find the ranges of the Wilson coefficients given in
\reftab{tab:NP:wc} using the $90\%$ C.L. constraints $\BRincl{e}{>0.04}{} < 6.8
\cdot 10^{-6}$ and $\BRincl{e}{>0.04}{}< 6.3 \cdot 10^{-6}$, see
\reftab{tab:comparison}.  As anticipated after \refeq{eq:MST},
$\BRincl{l}{>0.04}{}$ constrains $C^l_{T, T5}$ stronger than
$C^{l(\prime)}_{S,P}$.

The NP effects in $\FHe$ and $\FHmu$ are huge with respect to the SM predictions
as can be seen in \reftab{tab:NP:res}. $\FHl$ increases for increasing
$\BRincl{l}{1,6}{}$. This correlation is shown in the right-hand plot of
\reffig{fig:sc4} for contours of $\BRincl{\mu}{1,6}{} < \{1.75, 2.0, 2.25, 2.56
\} \cdot 10^{-6}$ in the $\FHmu - R_K$ plane. Similar correlations hold for the
electron sector.  $R_K$ receives order $10\%$ corrections from NP which are well
above the theoretical uncertainties. The branching ratios $\BRl$ are subject to
NP contributions $\lesssim +10 \%$, which cannot be separated from the larger
form factor induced uncertainties. On the other hand, the NP enhancements due to
$C^l_{T, T5}$ in $\BRincl{l}{1,6}{}$ are larger, about $70 \%$, which makes the
inclusive decays a sensitive probe of tensor operators.  As in all other
Scenarios I-III $\AFBe$ is negligibly small.  $|\AFBmu|$ does not exceed $3\%$.

%
%----------------------------------------------------
\subsection{Models with Scalar and Tensor Interactions \label{sec:tensors}}

While there are several known models beyond the SM with large (pseudo-) scalar
interactions, tensor operators are often neglected.  Let us begin with some
general remarks on the origin of $\bsll$ tensor operators in the SM and the
MSSM: In the SM they arise only at higher order in the electroweak operator
product expansion (OPE) from finite external momenta in the matching
calculation. In the MSSM tensor operators are induced at leading order OPE only
from photino and zino box diagrams, which are, however, subleading in $\tan
\beta$ with respect to the Higgs penguins discussed below.  Higgsino
contributions to tensors are further suppressed by down-type quark and lepton
Yukawa couplings \cite{christoph-privat}.  In addition tensors with two leptons
are induced by scalar operators under QED renormalization group running, hence
are of higher order in $\alpha_e/(4 \pi) \cdot \log(\mu/\mu_b)$, e.g.,
\cite{Hiller:2003js,Borzumati:1999qt}. Another mechanism to generate tensor
contributions is to consider models with scalars having appropriate quantum
numbers such that tree level exchange induces the operators $(\bar l_L b_R)
(\bar s_L l_R)$ or $(\bar l_R b_L) (\bar s_R l_L)$.  Subsequent fierzing then
leads to tensor operators. Among this class of models are those with
leptoquarks. We consider such models below after briefly commenting on the MSSM
at large $\tan \beta$ and the MSSM with broken $R$-parity.

For large values of $\tan \beta$ the MSSM produces substantial scalar couplings
$C_{S,P}^l$ from Higgs penguins, for example, induced by chargino loops
\begin{equation}
  C_{S,P}^l \propto \frac{m_l m_b}{m_{A^0}^2} \tan^3\beta , ~~~~~~
  C_{S,P}^{l \prime} \simeq \frac{m_s}{m_b} C_{S,P}^l ,
\end{equation}
for exact Wilson coefficients see \cite{Bobeth:2001sq}.  Here, $m_{A^0}$ denotes
the mass of the pseudoscalar Higgs boson. The relation $C_S^l = - C_P^l$ holds
only at leading order in $\tan\beta$ \cite{Bobeth:2001sq} and prevents the
generation of tensor couplings from QED running \cite{Hiller:2003js}. The
flipped coefficients $C_{S,P}^{l \prime}$ are suppressed by the mass of the
strange quark.  Since $C^l \propto m_l$ the couplings to electrons are
negligible.

In $R$-parity violating supersymmetry scalar and pseudoscalar FCNC-operators can
be generated at tree level from the superpotential (see, e.g.,
\cite{Grossman:1996qj})
\begin{equation}
  W_{ \not R} = \lambda_{ijk} L_L^i L_L^j \bar e_R^k + \lambda_{ijk}^\prime L^i_L Q^j_L \bar d^k_R ,
\end{equation}
where $L_L (Q_L)$ and $e_R (d_R)$ denote the superfields containing the lepton
(quark) doublet and the charged lepton (down-type quark) singlet, respectively.
One gets at the matching scale from sneutrino exchange
\begin{equation}
  \label{eq:RPV-C}
  C_{S,P}^l \propto 
  \frac{(4 \pi)^2 }{e^2}\frac{\lambda'^{\ast}_{k23}\lambda_{kll}}{V_{tb}^{} V_{ts}^* G_F m_{\tilde{\nu}_{k}}^2} , ~~~~~~
  C_{S,P}^{l  \prime} \propto \frac{(4 \pi)^2 }{e^2}
  \frac{\lambda'_{k32}\lambda^{\ast}_{kll}}{V_{tb}^{} V_{ts}^* G_F  m_{\tilde{\nu}_{k}}^2}.
\end{equation}
Here $ m_{\tilde{\nu}}$ denotes the sneutrino mass and summation over the
sneutrino flavor $k$ is understood. Contributions from squark exchange modify
only the vector and axial vector type operators and are not shown.  The
couplings in \refeq{eq:RPV-C} obey $C_S^l=-C_P^l$ and $C_S^{l\prime}=+C_P^{l
  \prime}$. As in the MSSM at large $\tan \beta$ with unbroken $R$-parity
discussed previously, there are no tensor operators generated from leading order
matching. The chirality flipped contributions can be sizeable and help to escape
the constraint from $\Bstoll$. The couplings \refeq{eq:RPV-C} are then
essentially only constrained by the $\bar B \to (K^{(*)},X_s) \bar e e$ and
$\bar B \to (K^{(*)},X_s) \bar \mu \mu$ branching ratios. The size of the
possible modification of $R_K$ from one is then given by the theoretical and
experimental uncertainties of the branching ratios, see also \cite{Xu:2006vk}.

The Lagrangian of leptoquarks $\phi_{LQ}$ coupling to a lepton and a quark can
be written as
\begin{align}
  {\cal {L}}_{LQ-l-q} = \lambda_{ij} l_i q_j \phi_{LQ} ,
\end{align}
where $i,j$ label the lepton and quark generation, respectively.  $\phi_{LQ}$
can be a scalar or a vector under space-time transformations and a singlet,
doublet or triplet under $SU(2)$.  Details can be seen, e.g., in
\cite{Davidson:1993qk}, where also contributions to flavor-changing scalar and
pseudoscalar operators have been discussed. Here we consider only the
contributions to tensor operators, which are induced by tree level scalar
leptoquark exchange and fierzing as explained earlier. The $SU(2)$-properties of
the requisite operators require mixing of leptoquarks with different
$SU(2)$-quantum numbers. The latter is induced by interactions with the Higgs
boson and arise after electroweak symmetry breaking, see the second reference in
\cite{Davidson:1993qk}. As a result, tensor operators in leptoquark models are
suppressed by the vacuum expectation value of the Higgs $v=<H^0>=(2 \sqrt{2}
G_F)^{-1/2}$ over the scalar leptoquark mass $m_S$.  Specifically for $b \to s
\bar l l$ transitions this yields the tensor coefficients
\begin{eqnarray}
  C^l_{T,T5} \propto  \frac{(4 \pi)^2 }{e^2} 
  \frac{ \lambda^*_{l3} \lambda_{2 l}}{V_{tb}^{} V_{ts}^* G_F  m_S^2} 
  \frac{v^2}{m_S^2} .
\end{eqnarray}

%
%
%----------------------------------------------------
\section{Summary \label{sec:conclusions}}

We thoroughly investigated the angular distributions in $\BtoKll$ decays in a
model-independent way. We find that the $\cos \theta$-dependence in the
normalized $1/\Gaml \, d\Gaml/d\!\cos\theta$ spectrum, see \refeq{eq:babar},
offers great opportunities to test the SM and search for NP. The requisite
observables are the flat term in the distribution, $\FHl/2$ and the
forward-backward asymmetry $\AFBl$. The coefficient of $\cos^2 \theta$ is
related to $\FHl$. No powers of $\cos \theta$ greater than two appear in the
$\BtoKll$ angular distribution up to higher dimensional operators not present in
${\cal{H}}_{\rm eff}$ \refeq{eq:Heff}-\refeq{eq:nonSM:op} and QED corrections.
Both are strongly suppressed by powers of the low energy masses and momenta over
the scale of electroweak NP and by $\alpha_e/(4 \pi)$, respectively.

In the SM, $\FHl \propto m_l^2$, and $\FHe$ is negligible.  The SM value for
$\FHmu$ is small, order few percent, and can be cleanly predicted using QCDF for
low dilepton masses with $2 \%$ accuracy, see \reftab{tab:SM:res}.  Taking into
account subleading $1/E$-corrections the uncertainty is conservatively inflated
to $\sim 6 \%$.  The forward-backward asymmetry vanishes exactly in the SM up to
the aforementioned higher order OPE and QED corrections.  The
$\alpha_e$-corrections induce the parametrically leading contribution of the
order $\alpha_e/(4 \pi)$.

We also give SM predictions for the $\BmtoKmmumu$ and $\B0toK0mumu$ branching
ratios. They have a substantial uncertainty of order $32 \%$ mostly from the
form factor.  On the other hand, the SM ratio of $\BtoKmumu$ to $\BtoKee$ decay
rates, $R_K^{\rm SM}$, equals one at the level of $10^{-4}$.  We show
analytically at large recoil using form factor symmetry relations that the
apparent huge suppression of lepton flavor effects in $R_K^{\rm SM}$ results
from the cancellation of ${\cal{O}}(m_l^2)$-corrections to leading order in
$1/E$ and $\alpha_s$ in the $\BtoKll$ decay rate, see \refeq{eq:Gaml:SM}.
In addition potentially large corrections to $R_K$ can arise from collinear
QED logarithms, whose actual net effect depends on experimental cuts
\cite{Huber:2005ig}. The corresponding calculation
for $\BtoKll$ decays has not been done.

Beyond the SM, the observables $\FHl$, $\AFBl$ and $R_K$ are sensitive to Higgs
and tensor interactions.  We work out NP signatures and correlations by taking
into account existing data on ${\cal{B}}(\Bstoll)$ and ${\cal{B}}(\bar B \to X_s
\bar l l)$ for $l=e$ and $l=\mu$ separately.  We find that the NP modifications
to the angular observables $\FHe$, $\FHmu$, $\AFBmu$ and $R_K-1$ can be
sizeable, see \reftab{tab:NP:res}.  Even larger effects in the forward-backward
asymmetries $\AFBmu$ and $\AFBe$ arise in models where both (pseudo-) scalar and
tensor operators are present. From a scan of twelve real NP coefficients
$C_{S,P}^{l(\prime)}, C^l_{T,T5}$ for $l=e$ and $l=\mu$ we find
model-independently the upper bounds
\begin{align}
  &  |\AFBe|  < 13  \%, &  &  |\AFBmu| < 15  \%.
\end{align}

Both $R_K$ and $\FHmu$ enable precision tests of the SM in exclusive $\BtoKll$
decays, but their experimental requirements are different: Whereas $R_K$
requires only measurements of decay rates into both electrons and muons, $\FHmu$
is extracted from the muon channel alone, however, at the price of an angular
analysis. The latter needs high statistics and is well suited for the LHC(b)
setup.  NP searches with angular distributions in $\BtoKmumu$ should also be
feasible at the Tevatron, where CDF has recently measured ${\cal{B}}(B^+ \to K^+
\bar \mu \mu)$ and ${\cal{B}}(B^0 \to K^{*0} \bar \mu \mu)$ \cite{Scuri:2007py}.

The experimental situation for the observables $\FHl$, $\AFBl$ and $R_K$ is
currently at a very early stage, see \reftab{tab:comparison}.  In particular,
all measurements average $l=e$ and $l=\mu$ final states except the ones of $R_K$
\cite{Abe:2004ir, Aubert:2006vb}. Ultimately all observations in rare
semileptonic decays $\BtoKll, \BtoKastll$ and $\BXsll$ should be available for
each lepton flavor separately since deviations from the SM could be
$l$-dependent. For example, NP in the electron channel could escape the
$\BtoKmumu$ decay studies completely implying also $R_K <1$.  Existing data on
the $l=e$ modes are weaker than the corresponding ones for decays into muons,
allowing for larger NP effects in the electron modes.  In this way, $b \to s
\bar e e$ induced channels such as $\BtoKee$ provide unique opportunities for
the clean $B$ factory environment.  Appropriate cuts in $q^2$ should be taken
into account to maximally exploit the theoretical predictions.

%
%
%----------------------------------------------------
\acknowledgments

We are happy to thank Yuehong Xie for stimulating questions and Thorsten
Feldmann and Uli Haisch for helpful communication.  
G.P.~is supported by a grant from the
G.I.F., the German-Israeli-Foundation for Scientific Research and Development.
The work of C.B.~is supported by the Bundesministerium f\"ur Bildung und
Forschung, Berlin-Bonn.  G.H.~gratefully acknowledges the hospitality and
stimulating atmosphere provided by the Aspen Center for Physics during the final
phase of this work.

%
%
%----------------------------------------------------
\appendix

%----------------------------------------------------
\section{$\BtoKll$ Form factors \label{app:formf}}

In this appendix we give definitions and properties of the heavy-to-light form
factors for the $\bar B \to K$ transition at large recoil. The symmetry
relations emerging in this region between the QCD form factors are reviewed
including symmetry breaking corrections.  Furthermore, details about the form
factor $f_+$ from Light Cone Sum Rules calculations \cite{Ball:2004ye} can be
found here.

The $\bar B \to K$ matrix elements are parametrized in terms of the three QCD
form factors $f_+, f_0$ and $f_T$ as \cite{Wirbel:1985ji}
\begin{align}
  \langle K(p_K) | \bar{s} \gamma_\mu b | \bar{B}(p_B) \rangle & =
    (2 p_B - q)_\mu f_+(q^2) + \frac{M_B^2 - M_K^2}{q^2} q_\mu [f_0(q^2) - f_+(q^2)],
\\
  \langle K(p_K) | \bar{s} i \sigma_{\mu\nu} q^\nu b |\bar{B}(p_B) \rangle & =
  - [(2 p_B - q)_\mu q^2 - (M_B^2 - M_K^2) q_\mu] \frac{f_T(q^2)}{M_B + M_K}.
\end{align}
At leading order in the $1/E$ expansion $f_{+,0,T}(q^2)$ obey symmetry relations
\cite{Charles:1998dr, Beneke:2000wa} such that they all can be related to a
single form factor denoted by $\xi_P(q^2)$. Within QCDF a factorization scheme
has been chosen with $f_+(q^2) \equiv \xi_P(q^2)$ \cite{Beneke:2000wa}.
Including subleading corrections, the symmetry relations can be written as
\begin{align}
  \nonumber
  \frac{f_0}{f_+} & = \frac{2 E}{M_B} \left[ 1 
    + \order{\alS} + \order{\frac{q^2}{M_B^2} \sqrt{\frac{\LamConf}{E}}} \right], 
\\[2mm]
  \label{eq:ff:sym:rel}
  \frac{f_T}{f_+} & = \frac{M_B + M_K}{M_B} \left[ 1
    + \order{\alS} + \order{\sqrt{\frac{\LamConf}{E}}} \right],
\end{align}
up to higher order QCD, power and mixed corrections.  The $\alS$-corrections
from the soft-overlap and hard scattering contributions indicated in
\refeq{eq:ff:sym:rel} have been calculated in QCDF and are given in
\cite{Beneke:2000wa}.  These corrections are taken into account in the numerical
analysis of this work.  Analogous relations can be found in the framework of
SCET using $\MSbar$ subtractions \cite{Bauer:2000yr,Hill:2004if}.
The symmetry relation breaking corrections due to subleading orders in the
$\LamConf/E$ expansion have been considered for the soft-overlap part using SCET
\cite{Beneke:2002ph}.  The corresponding corrections are indicated in
\refeq{eq:ff:sym:rel}.  Note that the expansion parameter is rather
$\sqrt{\LamConf/E}$ than $\LamConf/E$, and that for $f_0/f_+$ an additional
suppression of $q^2/M_B^2$ appears.
Subleading contributions from hard spectator scattering to \refeq{eq:ff:sym:rel}
are unknown and arise at higher order, ${\cal{O}}(\alpha_s \sqrt{\LamConf/E})$.

The form factor symmetry relations \refeq{eq:ff:sym:rel} imply in the SM for
$\BtoKll$ decays
\begin{equation}
   \label{eq:low:q2:rel}
   \frac{q^2}{M_B^2} |\tilde{F}_P|^2 + 4 |F_A|^2 + \frac{M_B^2 - M_K^2 + q^2}{M_B^2}
   2 Re(\tilde{F}_P F_A^\ast) = 
   \order{ \alS, \frac{q^2}{M_B^2} \sqrt{\frac{\LamConf}{E}}},
\end{equation}
which enters $a_l^{\rm SM}+c_l^{\rm SM}$, see \refsec{sec:dec:dist}.  Here, the
explicit SM expressions for $F_{V, A, P}$ \refeq{eq:Fis:def} have been used and
$F_P = m_l \tilde{F}_P$ has been rescaled.  The relation \refeq{eq:low:q2:rel}
involves only the ratio $f_0/f_+$ and results in a beneficial $q^2/M_B^2$
suppression of the power corrections.  Beyond the SM the corresponding
expression depends on all functions $F_i$ and there are in general no
cancellations from symmetry relations.

\begin{figure}
\begin{center}
  \epsfig{figure=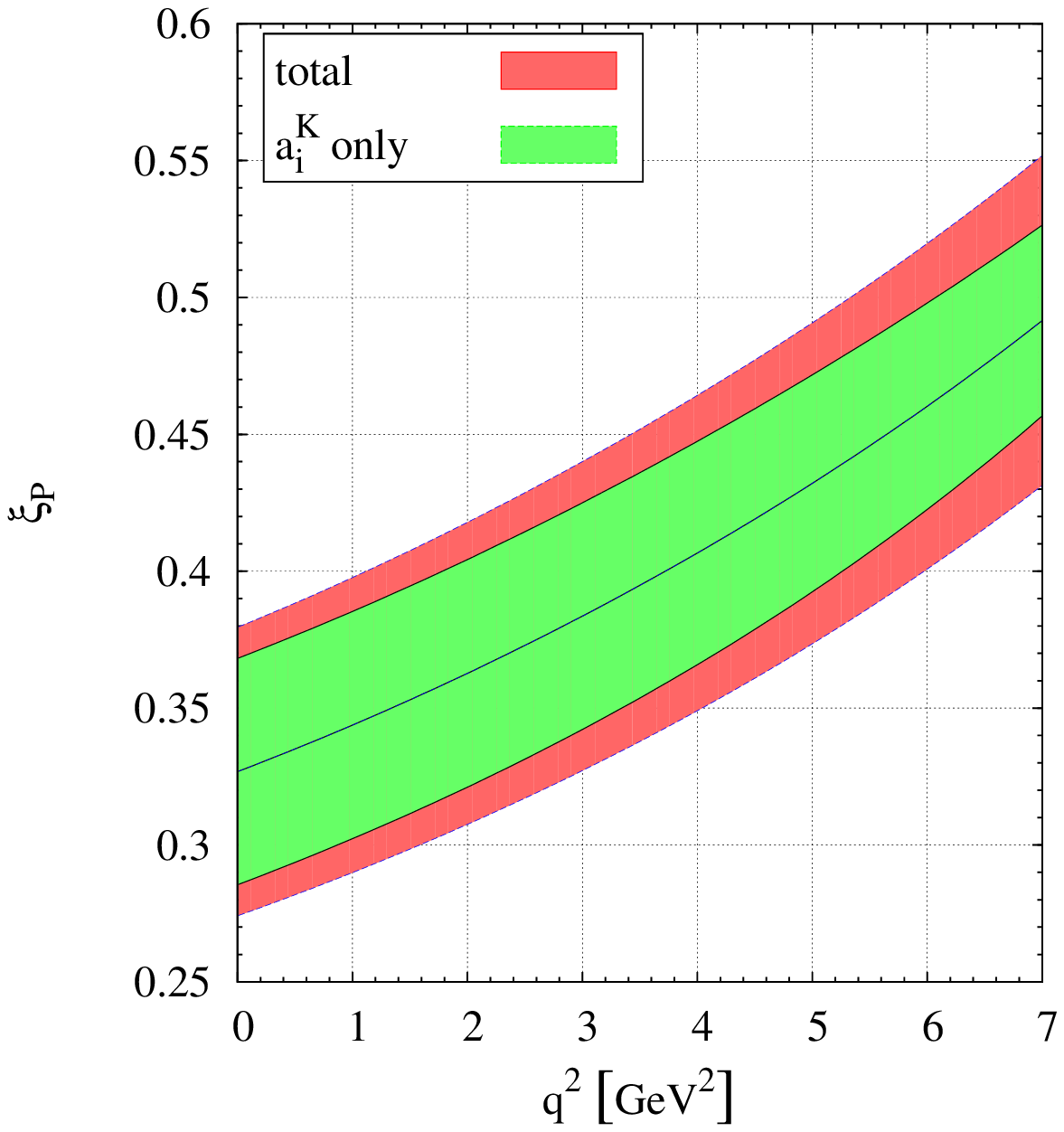,      height=7.5cm, angle=-0} 
  \epsfig{figure=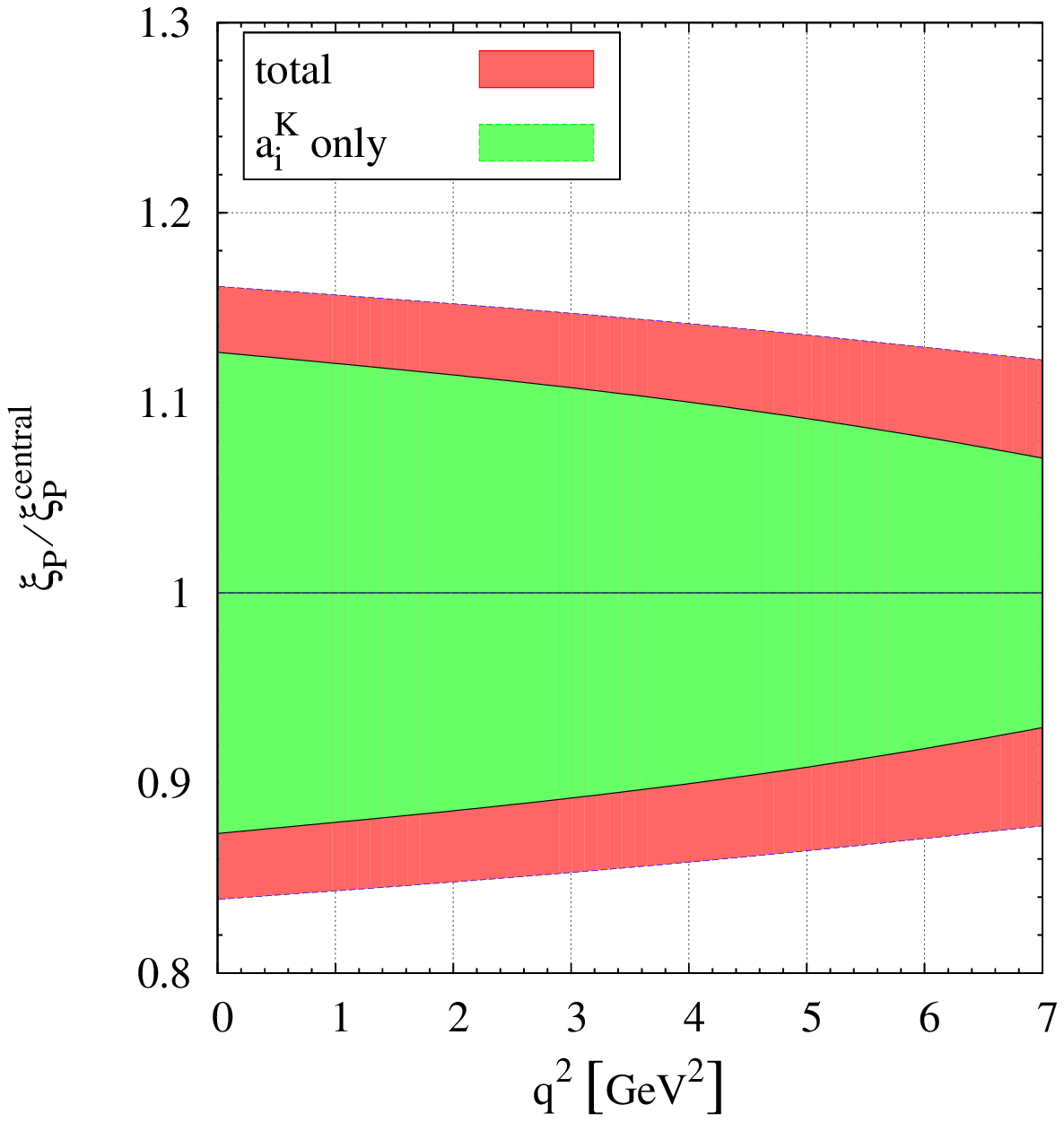, height=7.5cm, angle=-0}
\end{center}
\caption{ \label{fig:xi:q2} The form factor $\xi_P(q^2) = f_+(q^2)$ in the
  low-$q^2$ region including uncertainties from the Gegenbauer moments $a_i^K$
  (lighter shaded area) and from $a_i^K$ and $\Delta_{as}$ with their
  uncertainties added in quadrature (darker shaded area), for details see text.
  In the left-hand plot is shown $\xi_P(q^2)$, and in the right-hand plot the
  form factor normalized to its central value, $\xi_P(q^2)/\xi_P^{\rm
    central}(q^2)$.}
\end{figure}

We employ the form factor $f_+(q^2)=\xi_P(q^2)$ from LCSR calculations
\cite{Ball:2004ye}. It is given in terms of the Gegenbauer moments of the
$K$-meson LCDA, $a_1^K, a_2^K$ and $a_4^K$ as
\begin{equation}
  f_+(q^2) = f_+^{as}(q^2) + a^K_1(\mu_{IR}) f^{a_1}_+(q^2) 
    + a^K_2(\mu_{IR}) f^{a_2}_+(q^2) + a^K_4(\mu_{IR}) f^{a_4}_+(q^2).      
\end{equation}
The $q^2$-dependent functions $f_+^{a_i}$, $i = 1,2,4$ and $f_+^{as}$ are
obtained from a fit and parametrized in \cite{Ball:2004ye}. Here we use ``set
2'' with $m_b^{pole}= 4.8 \GeV$ corresponding to the infrared factorization
scale $\mu_{IR}=\sqrt{M_B^2-m_b^{pole \, 2}}=2.2 \GeV$. The running of the
Gegenbauer moments given in \reftab{tab:num:input} from $1 \GeV$ to $2.2 \GeV$
is accounted for by the scaling factors $\{ 0.793,0.696,0.590 \}$ for
$\{a^K_1,a^K_2,a^K_4 \}$.  The relative uncertainty of $f_+$ due to the
asymptotic form factor $f_+^{as}$ (which is independent of the $a_i^K$) at $q^2
= 0$ is approximately $\Delta_{as}/f_+(0)=10\%$, see Table~2 of
\cite{Ball:2004ye}.  In order to estimate the form factor uncertainty in the
low-$q^2$ region we scan over the Gegenbauer moments according to the ranges in
\reftab{tab:num:input} translated to $\mu_{IR}=2.2 \GeV$ and add the uncertainty
from $\Delta_{as}$ in quadrature. The form factor $f_+(q^2)=\xi_P(q^2)$ with its
uncertainties with and without $\Delta_{as}$ is shown in \reffig{fig:xi:q2}. The
total uncertainty is $16 \%$ at maximal recoil and reduces to $12 \%$ at $q^2 =
7 \GeV^2$. The reduction of the relative form factor uncertainty towards larger
values of $q^2$ stems from the increase of the form factor in this region while
keeping $\Delta_{as}$ from $q^2=0$. The decrease of the form factor uncertainty
for $q^2 >0$ has been considered likely in \cite{Ball:2004ye}.

%
%----------------------------------------------------
\section{${\cal T}_P$ for $\bar{B} \to K$ \label{app:calT}}

The amplitude ${\cal T}_P(q^2)$ can be extracted from \cite{Beneke:2001at} as
\begin{align}  \label{eq:app:calT}
  {\cal T}_P & = \xi_P \left[ C_P^{(0)} + \frac{\alS \CF}{4\pi} \left( C_P^{(\rm f)} +
    C_P^{(\rm nf)} \right)\right] \\
  & + \frac{\pi^2}{\Nc} \frac{f_B f_K}{M_B} \sum_\pm \int \frac{d\omega}{\omega}
      \Phi_{B,\pm}(\omega) \int_0^1 du\, \Phi_K(u) 
      \left[ T_{P,\pm}^{(0)} + \frac{\alS \CF}{4\pi} \left( T_{P,\pm}^{(\rm f)}
          + T_{P,\pm}^{(\rm nf)} \right)\right] (\omega,u), \nonumber
\end{align}
where all $ T_{P,\pm}^{(0)}, T_{P,\pm}^{(\rm f)}$ and $T_{P,\pm}^{(\rm nf)}$ are
functions of $(\omega,u)$.  $f_B$ and $f_K$ denote the $B$- and $K$-meson decay
constants, respectively, whereas $\Phi_{B,\pm}(\omega)$ and $\Phi_K(u)$ are the
corresponding LCDA's. The remaining quantities are calculable perturbatively
\begin{align}
  C_P^{(0)} & = - C_\parallel^{(0)}, & 
  C_P^{(\rm nf)} & = - C_\parallel^{(\rm nf)}, & 
  T_{P,\pm}^{(0)} & = - T_{\parallel,\pm}^{(0)}, &
  T_{P,\pm}^{(\rm nf)} & = - T_{\parallel,\pm}^{(\rm nf)}, 
  \nonumber
\end{align}
\begin{align}
  C_P^{(\rm f)} & = - C_7^{\rm eff} \left[ 4 \ln\frac{m_b^2}{\mu^2} + 2 L - 4 +
    4 \frac{\mu_f}{m_b} \right], & 
  T_{P,+}^{(\rm f)} & = - C_7^{\rm eff} \frac{4 M_B}{E (1-u)}, &
  T_{P,-}^{(\rm f)} & = 0.
  \label{eq:C:and:T}
\end{align}
Here the shift of the $b$-quark mass in $\Op_7$ from the $\MSbar$ to the PS
scheme has been taken into account in $C_P^{(\rm f)}$.  All quantities $L,
C_7^{\rm eff}$, $C_\parallel^{(0)}, C_\parallel^{(\rm nf)} $ and
$T_{\parallel,\pm}^{(0)}, T_{\parallel,\pm}^{(\rm nf)}$ in \refeq{eq:C:and:T}
are given in \cite{Beneke:2001at}.  The expressions for $C_P^{(\rm f)}$ and
$T_{P,+}^{(\rm f)}$ in \refeq{eq:C:and:T} agree with $- C_\parallel^{(\rm f)}$
and $- T_{\parallel,+}^{(\rm f)}$ given in \cite{Beneke:2004dp}, respectively,
where a different definition of the longitudinal form factor $\xi_\parallel$
with respect to the one used in previous works \cite{Beneke:2001at,
  Beneke:2000wa} is employed.

%
%----------------------------------------------------

\end{document}